\documentstyle[aps,12pt,epsf]{revtex}

\def\be{\begin{equation}}
\def\ee{\end{equation}}
\def\disp{\displaystyle}

\def\R{{\sf I\kern-.15em R}}
\def\E{{\sf I\kern-.15em E}}
\def\C{\kern.1em{\raise.47ex\hbox{$\scriptscriptstyle |$}}
             \kern-.40em{\sf C}}
\def\Z{{\sf Z\kern-.45em Z}}

\def\otm{\mbox{\tiny mark}}
\def\neot{\mbox{\tiny nonmark}}
\def\ver{\mbox{\tiny ver.}}
\def\gor{\mbox{\tiny hor.}}
\def\vse{\mbox{\tiny all}}

\newcounter{fig}

\begin{document}

\title {Thermodynamics and Topology of Disordered Systems: Statistics of the
Random Knot Diagrams on Finite Lattices}

\author{O. A. Vasilyev$^{\dag a}$ and S. K. Nechaev$^{\dag \ddag b}$}

\address{$^{\dag}$ Landau Institute of Theoretical Physics, Russian Academy of
Sciences, \\ Chernogolovka, Moscow region, 142432 Russia}

\address {$^{\ddag}$UMR 8626, CNRS--Universit\'e Paris XI, LPTMS, Universit\'e
Paris Sud, \\ 91405 Orsay Cedex, France}

\date{Journal of Experimental and Theoretical Physics, {\bf 93} (2001), 
1119--1136}

\maketitle

\vspace{0.5in}

\begin{abstract} 
The statistical properties of random lattice knots, the topology  of which is
determined by the algebraic topological Jones-Kauffman  invariants has been
studied by analytical and numerical methods. The Kauffman  polynomial invariant
of a random knot diagram has been represented by a  partition function of the
Potts model with a random configuration of ferro-  and antiferromagnetic bonds,
which allowed the probability distribution of  the random dense knots on a flat
square lattice over topological classes to be studied. A topological class is
characterized by the highest power of  the Kauffman polynomial invariant and
interpreted as the free energy of a $q$--component Potts spin system for $q
\to \infty $. It is shown that the highest power of  the Kauffman invariant is
correlated with the minimum energy of the  corresponding Potts spin system. The
probability of the lattice knot  distribution over topological classes was
studied by the method of transfer  matrices, depending on the type of local
crossings and the size of the flat  knot diagram. The obtained results are
compared to the probability  distribution of the minimum energy of a Potts
system with random ferro-  and antiferromagnetic bonds.  
\end{abstract}

\vspace{2.7in}

\hrule
\bigskip

$^{a}$ E-mail: vasilyev@itp.ac.ru

$^{b}$ E-mail: nechaev@ipno.in2p3.fr

\newpage

\tableofcontents

\section{Introduction}

New interesting problems are formulated, as a rule, in the boundary  regions
between traditional fields. This is clearly illustrated by the  statistical
physics of macromolecules, which arose due to the  interpenetration of the
solid state physics, statistical physics, and  biophysics. Another example of a
new, currently forming field is offered by  the statistical topology born due
to merging of the statistical physics,  theory of integrable systems, algebraic
topology, and group theory. The  scope of the statistical topology includes, on
the one hand, mathematical  problems involved in the construction of
topological invariants of knots and links based on some solvable models
and, on the other hand, the  physical problems related to determination of the
entropy of random knots  and links. In what follows, we pay attention
predominantly to problems  of the latter kind that can be rather conventionally
separated into a  subfield of "probabilistic-topological" problems\cite{Nech}.
Let us dwell on this  class of problems in more detail.

Consider a lattice embedded into a three-dimensional space, for which $\Omega$
is  the ensemble of all possible closed self-nonintersecting trajectories of
$N$  steps with a fixed point. Let  $\omega$ denote some particular realization
of such  a trajectory. The task is to calculate the probability  $P$
for a knot of specific topological class for all trajectories $\omega  \in 
\Omega$. This can be  formally expressed as
\be \label{eq:1}
P\{{\rm Inv}\}=\frac{1}{{\cal N} \left(\Omega\right)} \sum
\limits_{\{\omega \in \Omega\}} \delta \Big\{{\rm Inv}(\omega)-{\rm
Inv}\Big\}
\ee
where ${\rm Inv}(\omega)$ is a functional representation of the invariant for a
knot  on trajectory $\omega$, $\rm Inv$ is a particular topological invariant
characterizing  the given topological type of the knot or link, ${\cal
N}(\Omega)$ is the total number of  trajectories, and $\delta$ is the delta
function. The probability under  consideration should obey the usual
normalization condition $\sum \limits_{\{\rm Inv\}} P\{{\rm Inv}\}=1$. The
entropy $S$ of a knot of the given topological type is defined as
\be \label{eq:2}
S\{{\rm Inv}\}=\ln P\{{\rm Inv}\}
\ee
Based on the above definitions, it is easy to note that the 
probabilistic--topological problems are similar to those encountered in the 
physics of disordered systems and sometimes, as demonstrated below,  of
the thermodynamics of spin glasses. Indeed, the topological state plays  the
role of a "quenched disorder" and the functional $P\{{\rm Inv}\}$ is averaged
over  the trajectory fluctuations at a fixed "quenched topological state",
which is  similar to calculation of a partition function for a spin  system
with  "quenched disorder" in the coupling constant. In the context of this
analogy,  a question arises as to whether the concepts and methods developed
over  many years of investigation into the class of disordered statistical 
systems can be transferred to the class of probabilistic-topological  problems,
in particular, whether the concept of {\it self-averaging} is  applicable to
the knot entropy $S$.

The main difference between the systems with topological disorder and the 
standard spin systems with disorder in the coupling constant is a strongly 
nonlocal character in the first case: a topological state is determined  only
for the entire closed chain, is a "global" property of this chain, and  is
difficult to determine it for any arbitrary subsystem. Therefore, we may  speak
of the topology of a part of some closed chain only in a very rough 
approximation.  Nevertheless, below we consider a lattice model featuring a 
unique relationship between the topological disorder and the disorder in  the
local coupling constant for a certain disordered spin system on the  lattice.

Every time when we deal with topological problems, there arises the task  of
classification of the topological states. A traditional topological  invariant,
known as the Gauss invariant, in inapplicable because this  Abelian
(commutative) characteristic takes into account only a cumulative  effect of
the entanglement, not reflecting the fact that the topological state  depends
significantly on the sequence in which a given entanglement was formed. For 
example, when some trial trajectory entangles with two (or more)
obstacles,  there may appear configurations linked with several obstacles 
simultaneously, while being not linked to any one of these obstacles 
separately. In this context, it is clear that most evident questions 
concerning calculation of the probability of a knot formation as a result  of
the random closing of the ends of a given trajectory cannot be solved in  terms
of the Gauss invariant because this characteristic is incomplete.

A very useful method of the classification of knots was offered by a 
polynomial invariant introduced by Alexander in 1928. A breakthrough in  this
field took place in 1975-1976, when it was suggested to use the  Alexander
invariants for classification of the topological state of a  closed random
trajectory computer-simulated by the Monte Carlo method. The  results of these
investigations showed that the Alexander polynomials,  being a much stronger
invariants compared to the Gauss integral, offer a  convenient approach to the
numerical investigation of the thermodynamic  characteristics of random walks
with topological constraints. The  statistical-topological approach developed
in \cite{FrK} has proved to be very  fruitful: the main results gained by now
were obtained using this method  with subsequent modifications.

An alternative polynomial invariant for knots and links was suggested by  Jones
in \cite{Jones0,Jones}.  This invariant was defined based on the  investigation
of the topological properties of braids \cite{Birman,Prasolov}. Jones 
succeeded in finding a profound connection between the braid group 
relationships and the Yang-Baxter equations representing a necessary  condition
for commutativity of the transfer matrix \cite{Baxter}. It should be noted 
that neither the Alexander, Jones, and HOMFLY invariants, nor their various 
generalizations are complete. Nevertheless, these invariants are  successfully
employed in solving statistical problems. A clear geometric  meaning of the
Alexander and Jones invariants was provided by the results  of Kauffman, who
demonstrated that the Jones invariants are related to a partition function of
the Potts spin model  \cite{Kauffman}. Later Kauffman and Saleur showed that
the Alexander  invariants can be represented by a partition function in the
free fermion model \cite{Kauf_Sal}.

We employed the analogy between the Jones-Kauffman polynomial invariant  and
the partition function in the Potts model with ferro (f-) and 
antiferromagnetic (a-) links \cite{Wu0,Gr_Ne} to study the statistical
properties  of knots. In particular, the method of transfer matrices developed
in  \cite{BN,JC,SS} was used to determine the probability $P\{f_{K}\}$ of
finding a randomly  generated knot $K$ in a particular topological state
characterized by the  invariant $f_{K}$.

The idea of estimating $P\{f_{K}\}$ is briefly as follows: (i) the
Jones-Kauffman  invariant is represented as a partition function of the Potts
model with  disorder in the coupling constant; (ii) the thermodynamic
characteristic of  the ensemble of knots are calculated using the method of
transfer matrices  for the Potts spin system.

The paper is arranged as follows. Section 2 describes the model, defines  the
Jones-Kauffman polynomial invariant, and shows a connection of this  invariant
to the Potts model. Section 3 presents the numerical methods  employed and
introduces the necessary supplementary constructions. The  results of numerical
modeling are discussed and the conclusions are  formulated in Section 4.

\section{Knots on lattices: model and definitions}

Consider an ensemble of randomly generated dense knots on a lattice in a 
three-dimensional space. A knot is called "dense" if the string forming  this
knot is tightly fit to the lattice, not fluctuating in the space. In  this
case, the knots of various topological types possess no configuration  entropy
and the probability of formation of a knot belonging to a given  topological
type is determined only by a local topology of the system.

Not taking into account the fluctuational degrees of freedom of the  lattice
sites, this model is obviously very simplified. Nevertheless, this  approach
adequately reflects a physical situation, for example, in the  statistical
physics of condensed (globular) state of polymer  macromolecules. Virtually all
presently existing models taking into account  topological restrictions with
respect to the admissible spatial  configurations of polymer chains assume a
small density of the polymer,  that is, describe a situation far from the
compact state of globules. A  hypothesis of the essentially new globular phase
of a ring knotless  macromolecule (crumpled globule) existing at a large
polymer density (i.e.,  in the state of dense knots) was formulated in
\cite{GNSh}. Despite an indirect  experimental evidence for validity of the
crumpled globule hypothesis \cite{fr_gl},  direct observation of such objects
in real experiments or computer  simulations encounters considerable technical
difficulties. In this  context, investigation of the distribution of randomly
generated dense  knots of various topological types may provide for a better 
understanding of the structure of the phase space of knotted polymers in  the
globular state.

Analysis of a string configuration in the three-dimensional space is not  very
convenient for determining the topological type of a knot. The  standard method
consists in projecting the knot onto a plane in the general  position (with no
more than two knot segments intersecting at each point of  the plane) and
determining which segment line passes above (overcrossing) or  below
(undercrossing) in accordance with the knot topology. Such a  projection,
featuring over- and under-crossing, is called the diagram of a  knot. In what
follows we deal only with statistics of the knot diagrams.  Evidently, this
description implies an additional simplification of the  model, but we believe
that (in the phase of dense knots under  consideration) the additional
restrictions are not very significant since,  as noted above, the knot entropy
contains no contribution due to the string  fluctuations.

Thus, let us consider a square lattice of the size $N = L \times L$ on a
plane,  which is rotated for convenience by the $\pi /4$ angle. The lattice is
filled  by a (dense) trajectory featuring intersections at each lattice site
with  clear indications of the threads passing above and below (see an example
of  the $3 \times 3$ lattice in Fig.\ref{figpot}a). As can be readily verified,
a dense  trajectory on a lattice with an odd $L$ is unique, that is, represents
a knot  with the topology unambiguously determined by the pattern of over- and 
undercrossing and the boundary conditions. In this case, the probability of 
realization of the knot belonging to a given topological type is determined  by
the distribution of over- and undercrossing in all lattice sites. The  problem
considered in this study is to describe the distribution of dense  knots over
topological classes for various kinds of the over- and  undercrossing
distributions.

\subsection{Reidemeister moves and definition of the kauffman invariant}

In solving any topological problem, the main point consists in comparing  the
knots. A knot diagram on the plane obey the following Reidemeister 
theorem\cite{Radem}:{\it Two knots in the tree-dimensional space can be
continuously  transformed into each other if and only if the diagram of one
knot can be  transformed into that of the other knot by a sequence of local 
transformations (moves) of types I, II, and III} (see Fig.\ref{fig:radem}).

As can be seen in Fig.\ref{fig:radem}, the Reidemeister move I leads to the
formation  of a singularity on the plane during continuous fastening of the
loop; this  move is forbidden for smooth trajectories on the plane. Two knots
are  called regularly isotopic if their plane diagrams can be transformed into 
each other by means of the Reidemeister moves II and III. When the mutual 
transformation of the knot diagrams requires using the Reidemeister moves  of
all three types, the knots are called ambiently isotopic.

Consider a two-dimensional knot diagram as a graph in which all  intersection
points (vertices) are characterized by the order of crossing  (over- and
under-crossing). Then each intersection point belongs to one of  the two
possible crossing modes. Let a $k$th point of the graph be  characterized by
the variable $\epsilon_{k}$  acquiring the values $\pm 1$, depending on the 
mode of crossing.

Let us define the algebraic Kauffman invariant as a sum over all possible 
variants of splitting the diagram at the vertices. According to this, each 
splitting is ascribed a certain statistical weight by the following rule: a  
vertex with $\epsilon=+1$ is given the weight $A$ or $B$ corresponding to the
horizontal  or vertical splittings, respectively; for a vertex with
$\epsilon=-1$, the weight $B$  corresponds to the horizontal splitting and the
weight $A$, to the vertical splitting.  This definition can be illustrated by
the following scheme:
\be \label{xxx}
\unitlength=1.00mm
\linethickness{0.4pt}
\begin{picture}(134.00,31.00)
\put(12.00,16.00){\line(1,-1){4.00}}
\put(18.00,10.00){\line(1,-1){4.00}}
\put(22.00,16.00){\line(-1,-1){10.00}}
\put(27.00,14.00){\vector(2,1){6.00}}
\put(27.00,8.00){\vector(2,-1){6.00}}
\put(1.00,11.00){\makebox(0,0)[cc]{$\epsilon=+1$}}
\put(57.00,20.00){\makebox(0,0)[cc]{$A$}}
\put(57.00,3.00){\makebox(0,0)[cc]{$B$}}
\put(104.00,14.00){\vector(2,1){6.00}}
\put(104.00,8.00){\vector(2,-1){6.00}}
\put(79.00,11.00){\makebox(0,0)[cc]{$\epsilon=-1$}}
\put(134.00,20.00){\makebox(0,0)[cc]{$B$}}
\put(134.00,3.00){\makebox(0,0)[cc]{$A$}}
\put(90.00,16.00){\line(1,-1){10.00}}
\put(100.00,16.00){\line(-1,-1){4.00}}
\put(94.00,10.00){\line(-1,-1){4.00}}
\put(51.00,3.00){\oval(10.00,10.00)[l]}
\put(39.00,3.00){\oval(10.00,10.00)[r]}
\put(45.00,14.00){\oval(10.00,10.00)[t]}
\put(45.00,26.00){\oval(10.00,10.00)[b]}
\put(129.00,3.00){\oval(10.00,10.00)[l]}
\put(117.00,3.00){\oval(10.00,10.00)[r]}
\put(123.00,14.00){\oval(10.00,10.00)[t]}
\put(123.00,26.00){\oval(10.00,10.00)[b]}
\end{picture}
\ee
Thus, there are $2^N$ various microstates for the diagram of a knot  possessing
$N$ vertices. Each state $\omega$ of the knot diagram represents a set of 
nonintersecting and self-nonintersectiong cycles. The manifold of all
microstates is  denoted by $\{\omega\}$ (below, the braces $\{\dots \}$ will
denote summing over these  states). Let $S(\omega)$  be the number of cycles
for the microstate $\omega$. Consider  a partition function
\be \label{eqk}
f_{KR}= \sum \limits_{\{\omega\}}
d^{S(\omega)-1}A^{l(\omega)}B^{N-l(\omega)}
\ee
where the sum is taken over all $2^N$ possible splittings of the diagram;
$l(\omega)$  and $N-l(\omega)$  are the numbers of vertices with the weights
$A$ and $B$,  respectively, for a given set of splittings of the microstate
$\omega$. Kauffman  derived the following statement\cite{Kauffman}: a
polynomial $f_{KR}$ of the variables $A$,  $B$, and $d$ representing a
partition function (\ref{eqk}) is a topological invariant  of the regularly
isotopic knots, if and only if the parameters $A$, $B$, and $d$  obey the
relationships: $A B = 1$ and $A B d +A^{2}+B^{2}=0$. A proof of this statement
in \cite{Kauffman} is based on verification of the  invariance of the partition
function $f_{KR}$ with respect to the Reidemeister  moves II and III. (The
invariant for all tree Reidemeister moves is defined  below.) The latter
relationships pose the following restrictions on the parameters $A$, $B$, and
$d$ in Eq. (\ref{eqk}):
\be \label{eqab}
\begin{array}{l}
B=A^{-1}, \\ d=- A^{2}-A^{-2}
\end{array}
\ee
which imply that invariant (\ref{eqk}) is a polynomial of the single variable
$A$.

\subsection{A partition function of the Potts model as a bichromatic
polynomial}

Consider an arbitrary flat graph containing $N$ vertices. Let $i$th 
vertex be characterized by a spin variable  $\sigma_{i}$ ($1\le \sigma_{i}\le
q$) and each edge of the  graph, connect ing the $i$th and $j$th spins ($1\le
\{i,j\} \le N$), by the coupling  constant $J_{i, j}$ . The energy of the Potts
model is defied as \cite{Baxter}
$$ 
E=-\sum \limits_{\{i,j\}} J_{i,j} \delta(\sigma_{i},\sigma_{j})
$$
where the sum over $\{i, j\}$ is taken only for the adjacent spins connected 
by edges of the graph.  Then, the partition function can be expressed as
$$
Z=\sum \limits_{\{\sigma \}} \exp \left( \sum \limits_{\{i,j\}}
\frac{J_{i,j}}{T} \delta(\sigma_{i} , \sigma_{j}) \right)
$$
where $\{ \}$ denotes summation over all possible spin states, and the sum 
over $\{i, j\}$ is taken as indicated above. The last expression can be
written  in the following form:
\be \label{eqpp}
Z=\sum \limits_{\{\sigma \}}\prod \limits_{\{i,j\}} \left(1+v_{i,j}   
\delta(\sigma_{i},\sigma_{j}) \right),\;\;\;v_{i,j}=\exp
\left(\frac{J_{i,j}}{T} \right)-1
\ee
A pair of adjacent spins $i$ and $j$ introduces into the product term a 
contribution equal to $\exp(\frac{J_{i,j}}{T})$ for $\sigma_{i}=\sigma_{j}$ and
a unity contribution for  $\sigma_{i} \ne \sigma_{j}$. Let us perform a
procedure according to the following rules to a  given spin configuration and
the corresponding graph:
\begin{itemize}
\item an edge is  removed from the graph if a contribution to the above product
from the  spins connected by this edge is unity; 
\item an edge is retained in the graph  if the contribution from the spins
$\sigma_{i}$ and $\sigma_{j}$ connected by this edge is
$\exp(\frac{J_{i,j}}{T})$ . This procedure ensures a one-to-one mapping between
the spin  configuration corresponding to a product term in the sum
(\ref{eqpp})  and the  related set of graph components.
\end{itemize}

Consider a graph $G$ containing $M$ edges and $C$ connected components (an 
isolated vertex is considered as a separate component). Upon summing over 
all possible spin configurations and the corresponding subdivisions of the 
graph $G$, we may present the sum (\ref{eqpp}) in the following form:
\be \label{eqpg}
Z=\sum \limits_{\{G\}} q^{C} \prod \limits_{\{i,j\}}^{M} v_{i,j}
\ee
where $\{G\}$ denotes summing over all graphs and  $\prod   
\limits_{\{i,j\}}^{M}$ is the product over all  edges of the graph $G$.  Note
that expression (\ref{eqpg}) can be considered as an  analytic continuation of
the Potts spin system to non-integer and even  complex $q$ values.

For $v_{i,j}\equiv v$, expression  (\ref{eqpg}) coincides with a well-known
representation of  the partition function of the Potts model in the form of a
bichromatic  polynomial (see, e.g.,  \cite{Baxter,Wu})). The same expression is
involved in the  correspondence between the Potts model and the model of
correlated  percolation over sites and bonds suggested by Fortuin and
Kastelleyn \cite{FK},  which serves a base for the Monte Carlo cluster
algorithms developed by  Swedsen and Wang \cite{SW} and Wolff \cite{Wolff} for
the Potts model.

\subsection{Kauffman invariant represented as a partition function for the
Potts  model}

The Kauffman invariant of a given knot can be represented as a partition 
function for the Potts model on a graph corresponding to an arbitrary plane 
diagram of this knot, but we restrict the consideration below to an  analysis
of the knots on lattices.

Let us rewrite the Kauffman invariant in the form of a partition function  for
the Potts model determined in the preceding section, with $M$ denoting  the
lattice knot diagram (see Fig.\ref{figpot}a). The auxiliary variables 
$s_{k}=\pm 1$ 
describe the mode of the knot splitting in each $k$th lattice site, 
irrespective of the values of variables $\epsilon_{k}=\pm 1$  in the same
vertices:
\begin{center}
\unitlength=1.00mm
\linethickness{0.4pt}
\begin{picture}(86.00,16.00)
\put(71.00,6.00){\oval(10.00,10.00)[l]}
\put(59.00,6.00){\oval(10.00,10.00)[r]}
\put(5.00,11.00){\oval(10.00,10.00)[b]}
\put(5.00,-1.00){\oval(10.00,10.00)[t]}
\put(25.00,6.00){\makebox(0,0)[cc]{$s_k=+1$}}
\put(86.00,6.00){\makebox(0,0)[cc]{$s_k=-1$}}
\end{picture}
\end{center}

Let $\omega=\{s_{1}, s_{2}, \dots,s_{N} \}$ be the set of all variables
characterizing the  mode of splitting in the lattice containing $N$
intersections. The Kauffman  invariant (\ref{eqk}) can be written in the
following form:
\be \label{eqkeps}
f_{KR}(\epsilon_1,...,\epsilon_N)=\sum \limits_{\{\omega\}}
(-A^{2}-A^{-2})^{S(\omega)-1} \exp \left( \ln A \sum \limits_{k=1}^{N}
\epsilon_{k} s_{k}  \right)
\ee
Here, $\{\omega\}$ indicates summing over all values of the variables $s_{k}$
(i.e. over  all modes of splitting thy lattice diagram $M$) and the variables
$\epsilon_{k}$  characterize a particular realization of "quenched" disorder 
in the system.
Now we can demonstrate that configurations obtained as a result of splitting
the diagram  are in a one-to-one correspondence to configurations of the Potts
model on  a dual lattice.

Consider the Potts model lattice  $\Lambda$ corresponding to the lattice $M$
(see  Fig.\ref{figpot}b) , where open circles indicate positions of the Potts
spins). The $b_{i,j}$  edge of lattice $\Lambda$ corresponds to the $k$th site
of the lattice diagram $M$.  Therefore, disorder in the vertices of diagram $M$
set by the variables $\epsilon_{k}$ coincides with the disorder in edges
$b_{i,j}$ of the lattice $\Lambda$, that is,  with the disorder in coupling
constants. Let us define the disorder of  $b_{i,j}$ for the lattice $\Lambda$
via the coupling constants at the corresponding $k$th  site of the lattice
diagram $M$:
\be \label{eqbb}
b_{i,j}= \left\{
\begin{array}{cl}
-\epsilon_{k}, & \mbox{if the bond ($i,j$) is vertical} \\
\epsilon_{k}, & \mbox{if the bond ($i,j$) is horizontal}
\end{array} \right.
\ee

It should be recalled that the definition of the Kauffman invariant (\ref{eqk})
is based on splitting the lattice diagram $M$ into polygons representing a 
system of closed densely packed nonintersecting contours (Fig.\ref{figpot}b).
For a  given configuration of splitting the lattice diagram $M$ and the
corresponding  dual lattice $\Lambda$, we take the following agreement: all
edges of the lattice $\Lambda$ not crossed by polygons of the lattice $M$ are
labeled. In Fig.\ref{figpot}b, the  labeled edges are indicated by dashed
lines. All the remaining edges of  $\Lambda$ are unlabeled. In these terms, the
exponent in the partition function in Eq. (\ref{eqkeps}) can be  written as 
follows:
\be \label{eqsb}
\begin{array}{c}
\sum \limits_{k} s_{k} \epsilon_{k}=
\sum \limits_{\otm} s_{k} \epsilon_{k} + \sum \limits_{\neot} s_{k}
\epsilon_{k}=
\sum \limits_{\otm}^{\gor} s_{k} \epsilon_{k} +
\sum \limits_{\otm}^{\ver} s_{k} \epsilon_{k}+
\sum \limits_{\neot}^{\gor} s_{k} \epsilon_{k} +
\sum \limits_{\neot}^{\ver} s_{k} \epsilon_{k}= \\
\sum \limits_{\otm}^{\gor} b_{i,j}+
\sum \limits_{\otm}^{\ver} b_{i,j}-
\sum \limits_{\neot}^{\gor} b_{i,j}-
\sum \limits_{\neot}^{\ver} b_{i,j}=
-\sum \limits_{\vse} b_{i,j} +2 \sum \limits_{\otm}b_{i,j} \\
\end{array}
\ee
where
$$
\sum \limits_{\otm} b_{i,j}+ \sum \limits_{\neot} b_{i,j}=
\sum \limits_{\vse} b_{i,j}
$$
Let $m_{\omega}$ be the number of labeled edges and  $C_{\omega}$ be the
number of connected  components in the labeled graph  $\omega$  with $N_p$
vertices (each vertex  corresponding to a Potts spin). The Euler relationship
for this graph is
$$
S(\omega)=2C_{\omega}+m_{\omega}-N_{p}
$$
Now we can readily transform Eq. (\ref{eqkeps}) to
\be \label{eqkg}
f_{KR}(A,\{b_{i,j}\})=(-A^{2}-A^{-2})^{-(N_{p}+1)}\prod
\limits_{\vse}^{N} \left( A^{-b_{i,j}} \right) \sum
\limits_{\{G\}} \left(-A^{2}-A^{-2} \right)^{2C_{\omega}} \prod
\limits_{\otm}^{m_{\omega}}\left(A^{2 b_{k,l}} \left(-A^{2}-A^{-2}
\right) \right)
\ee
by using  (\ref{eqsb}) and the fact that $N$ is odd. Comparing 
expressions (\ref{eqkg}) and (\ref{eqpp}), we obtain the equality
\be \label{eqkp1}
\sum \limits_{\{G\}} \left(A^{2}+A^{-2} \right)^{2C_{\omega}}
\prod \limits_{\otm}^{m_{\omega}} \left(A^{2 b_{i,j}}
\left(-A^{2}-A^{-2}\right)\right) \equiv \sum_{\sigma} \prod_{\{i,j \}}
\left( 1+v_{i,j} \delta(\sigma_{i}, \sigma_{j})\right)
\ee
in which the right-hand part coincides with a partition  function of the 
Potts model represented in the form of a bichromatic polynomial. From this 
we obtain
$$
\begin{array}{l}
v_{i,j}=A^{2b_{i,j}}\left(-A^{2}-A^{-2} \right)=-1-A^{4 b_{i,j}} \\
q=(A^{2}+A^{-2})^2
\end{array}
$$
Since the disorder constants may acquire only discrete values $b_{i,j}=\pm 1$, 
we may write the following expression for the coupling constants 
$J_{i,j}$:
\be \label{eqj}
J_{i,j}=T \ln \left(1- \left(A^{2}+A^{-2} \right)A^{2b_{i,j}} \right)
= T \ln \left(-A^{4 b_{i,j}} \right)
\ee

Thus, we arrive at the following statement. The topological Kauffman  invariant
$f_{KR}(A)$ of the regularly isotopic knots on the lattice $M$  can be 
represented in the form of a partition function for a two-dimensional Potts 
model on the corresponding dual lattice $\Lambda$:
\be \label{eqkp2}
f_{KR}(A,\{b_{i,j}\})=K \left(A,\{b_{i,j} \} \right)\,
Z\left(q(A),\{J_{i,j}(b_{i,j},A) \} \right)
\ee
where
\be \label{eqh}
K\left(A,\{b_{i,j}\}\right) = \left(A^{2}+A^{-2}\right)^{-(N_{p}+1)}     
\exp\left(-\ln A \sum\limits_{\{ i,j\}} b_{i,j}\right)
\ee
is a trivial factor independent of the Potts spins. Here, the partition 
function of the Potts model is
\be \label{eqpb}
Z\left(q(A),\{J_{i,j}(b_{i,j},A) \} \right)=\sum \limits_{\{\sigma\}}
\exp \left(\sum \limits_{\{i,j\}}\frac{J_{i,j}(b_{i,j},A)}{T}
\delta(\sigma_{i},\sigma_{j})\right)
\ee
with the coupling constant $J_{i,j}$ and the number of states $q$ given by the 
formulas
\be \label{eqjq}
\frac{J_{i,j}}{T} = \ln (-A^{4b_{i,j}}),\qquad q=(A^{2}+A^{-2})^2
\ee
and the variable $b_{i,j}$ expressing disorder on edges of the lattice
$\Lambda$  corresponding to the lattice $M$. A relationship between $b_{i,j}$ 
and $\epsilon_{k}$ is  defined in Eq.  (\ref{eqbb}).

A specific feature of the partition function  (\ref{eqpb}) is the existence of
a  relationship between the temperature $T$ and the number of spin states $q$.
For  this reason, $T$ and $q$ cannot be considered as independent variables.
Once a  positive $q$ value is fixed, the variable $A$ can formally acquire the
complex  values according to Eqs. (\ref{eqjq}), where logarithm can be taken of
a complex  argument. The appearance of complex quantities in the partition
function  can be interpreted in two ways. On the one hand, this implies
expansion of  the domain of the partition function to the complex plane. On the
other  hand, the parameters $T$ and $J_{i,j}$ do not enter explicitly into an
expression  for the Kauffman invariant and, hence, their complex values do not
require  any special consideration. Below we will be interested mostly in the 
probability distribution of the maximum power in the polynomial invariant  and,
therefore, can digress from particular values of the variables $A$, $T$,  and
$J_{i,j}$.

Thus, we have determined an invariant $f_{KR}$ for the regularly isotopic 
knots, the diagrams of which are invariant with respect to Reidemeister  moves
II and II.  In order to obtain an invariant $f_{KR}$ for the ambiently
isotopic  knots with oriented diagram, the corresponding partition function has
to be  invariant with respect to the Reidemeister moves of all types. Let us 
characterize each oriented intersection by a variable $c_{k}=\pm 1$ according
to  the following rule

\vspace{0.4in}
\hspace{2cm}
\begin{picture}(110.00,10.00)
\linethickness{0.4pt}
\unitlength=0.89mm
\put(5.00,5.00){\makebox(0,0)[cc]{(a)}}
\put(12.00,0.00){\line(1,1){4.00}}
\put(18.00,6.00){\vector(1,1){4.00}}
\put(22.00,0.00){\vector(-1,1){10.00}}
\put(30.00,5.00){\makebox(0,0)[cc]{$c_{k}=-1$}}
\put(78.00,5.00){\makebox(0,0)[cc]{(b)¡}}
\put(85.00,0.00){\vector(1,1){10.00}}
\put(95.00,0.00){\line(-1,1){4.00}}
\put(89.00,6.00){\vector(-1,1){4.00}}
\put(103.00,5.00){\makebox(0,0)[cc]{$c_{k}=+1$}}
\end{picture}
\vspace{.1in} 

In addition, we define the knot twisting $Tw(\omega)$ as a sum of the
$c_{k}=b_{k}$ values  taken over all intersections: $Tw(\omega)=\sum
\limits_{k} c_{k}$ The invariant  $f_{KI}(\omega)$ of the ambiently isotopic
knots can be expressed as follows \cite{Kauffman}:
\be \label{eqtw}
f_{KI}(\omega)=f_{KR}(\omega) (-A)^{3Tw(\omega)}
\ee

It should be noted that our boundary conditions imply that $c_{k}=b_{k}$. As
is  known, the Kauffman polynomial invariant $f_{KI}(\omega;A)$ of the
ambiently isotopic knots as a  function of variable $A$ is equivalent to the
Jones polynomial invariant $f_{J}(\omega;x)$  of the variable $x=A^{4}$. Now we
can use formulas (\ref{eqkp2}) and (\ref{eqh}) to  express the Jones invariant
through a partition function of the Potts  model. Let  a partition function of
the Potts model have the form $Z(t;q)=\sum \limits_{E} H(E,q) t^{-E}$ where
$t=e^E$; $E$ denotes the energy levels over which the sum is taken, and 
$H(E,q)$ is the degree of degeneracy of an energy level $E$ for a given $q$ 
value. Taking into account that $x=A^{4}=-t$ and using formulas (\ref{eqkp2}),
(\ref{eqh}) and (\ref{eqtw}), we obtain the following expression  for the Jones
polynomial invariant:
\be \label{eqjn}
f_{J}
=\sum \limits_{E} H(E;q=2+x+x^{-1}) (-x)^{-E}(1+x)^{-N_{p}-1}
(-\sqrt{x})^{N_{p}+1+\sum \limits_{\{i,j\}}b_{i,j}}
\ee
In what follows, by the maximum power of a polynomial invariant we imply  the
maximum power of variable $x$ in the Jones polynomial invariant $f_{J}$. It 
will be born in mind that the power of the Kauffman polynomial invariant 
$f_{KI}$ of the ambiently isotopic knots is obtained from the corresponding
power of the  Jones polynomial invariant through simply multiplying by a factor
of four.

As indicated above, out task is to calculate the probability $P\{f_{J}\}$  of 
finding a knot on a lattice in the topological state with a preset J
ones-Kauffman 
invariant $f_{J}(x,\{\epsilon_{k}\})$ among all the $2^N$ possible disorder
realizations $\{\epsilon_{k}\},\;k=1,\dots,N$ This probability can formally be
written as 
$$   
P_N\{f_{J}\}=\frac{1}{2^N} \sum \limits_{\{\epsilon_{k}\}}   
\delta \Big(f_{J}(x,\{\epsilon_{1}, \dots,\epsilon_{N}\})-f_{J}\Big)
$$

Thus, the  topological disorder determined by a random independent selection
of  intersections of the $\epsilon=+1$ and $\epsilon=-1$ types represents a
random quenched  external field.  When we use the Jones invariant, each
topological class (homotopic type) is characterized by a polynomial.  In this
case,  precise identification of the homotopic type for a knot on a lattice 
containing $N$ intersections will require $N$ variables. Since the number of 
various homotopic types increases as $2^N$ , it is very difficult to study the
probability of each separate homotopic type characterized by $N$  variables.
For this reason, we introduce a simplified characteristic  of a knot on a
lattice - the maximum power $n$ of the polynomial invariant $f_{J}(x)$:
\be \label{eq:exp}
n=\lim_{|x|\to\infty}\frac{\ln f_{J}(x)}{\ln (x)}
\ee

For a trivial knot $n = 0$ and, as the knot complexity grows, the maximum 
power increases (not exceeding $N$). Thus, ensemble of all   of knots on a 
lattice can be divided into subclasses characterized by the power  $n$ ($0\le n
\le N$) of the Jones polynomial invariant. In these terms, we will study the 
probability that a randomly selected knot belongs to one of these  subclasses
(and is characterized by the maximum power $n$ of the Jones invariant).  Random
knots can be generated by two methods:
\begin{enumerate}
\item place a fixed number of intersections with $\epsilon=-1$ on the knot
diagram  with $N$ vertices (accordingly, the other vertices belong to
intersections of  the $\epsilon=+1$ type); 
\item place intersections of the  $\epsilon=-1$ and  $\epsilon_{k}=-1$ types in
each vertex with the probabilities $p$ and $1 - p$, respectively.
\end{enumerate}
As can be  readily seen, a trivial knot having all intersections of the
$\epsilon=+1$ type  corresponds to a partition function with the ferro (f-)
and  antiferromagmetic (a-) bonds distributed in accordance with the rule
(\ref{eqbb}).  An impurity (corresponding to $\epsilon_{k}=-1$ of the knot
diagram) will be considered as a change in the sign of $b_{i,j}$ relative to
the values  characterizing trivial knots. Note that we must differentiate
between the  notion of impurity (a change in the sign of $b_{i,j}$) from the
a-bond with  $b_{i,j}=-1$.  For a trivial knot with all $\epsilon_{k}=1$, the
lattice contains no impurities while containing the a-bonds.

\section{Auxiliary constructions and numerical methods}

\subsection{The form of a lattice for the Potts model and the positions of
ferro-  and antiferromagnetic bonds}

Now we describe the geometry of a lattice for the Potts model 
corresponding to a knot diagram of the $N = L \times L$ size. This study is 
restricted to square lattices, although all considerations remain valid for 
rectangular lattices as well. Figure~\ref{fig1}a  shows an example of the
trivial knot on a lattice with $N = 5 \times 5$. Positions of the Potts spins 
corresponding to this lattice diagram are indicated by circles. The Potts  spin
lattice corresponding to this knot is depicted in Fig.\ref{fig1}b, where the 
f-bonds (that would be horizontal in Fig.\ref{figpot}) are indicated by solid
lines  and the a-bonds (that would be vertical in Fig.\ref{figpot} ) are
indicated by dashed  lines. In Fig.\ref{fig1}c, the same lattice is transformed
into a rectangular one of  the $L_{h}\times L_{v}$ type, where $L_{h}=L+1$ and 
$L_{v}=(L+1)/2$ (in this  particular case, $L_{v}=3, \;\;\;L_{h}=6$. The
partition function of the Potts model is studied below for a rectangular
lattice of this type.

The fact that the height of the lattice for the Potts model is half that of 
the lattice knot diagram (see Figs.\ref{fig1}a--c) is very convenient in the
case  of using the method of transfer matrices, where the computational time 
expenditure exponentially depends on the lattice height. Let $N = L \times L$
be  the total number of bonds in the lattice, with $N_{+}=\sum
\limits_{\{i,j\}} \delta(b_{i,j},1)$ representing the number of f-bonds and
$N_{-}=\sum \limits_{\{i,j\}} \delta(b_{i,j},-1)$ the number of a-bonds. An
impurity, corresponding to $\epsilon_{k}=-1$ in the initial  lattice, is
described by a change in sign of the corresponding constant   $b_{i,j}$ for the
Potts model lattice.

It should be emphasized once again that by impurity we imply a change in  the
type of intersection from $\epsilon_{k}=+1$ to $\epsilon_{k}=-1$, which
corresponds to a  change in sign of the coupling constant $b_{i,j}$ for this
intersection, rather than to the a-bond (such bonds are present in the Potts
model lattice of  the trivial knot). A trivial knot is characterized by the
absence of  impurities on the Potts model lattice  $N_{+}=\frac{N+1}{2},\;
N_{-}=\frac{N-1}{2}$ Note that the Kauffman invariants for a knot and its
mirror image are  different. This is due to the following property of the Jones
invariant:  the invariant of the mirror image of a knot is obtained by
substituting  $t \to t^{-1}$ (and $A\to A^{-1}$), after which the distribution
of powers  in the  polynomial is asymmetric with respect to the substitution $p
\to 1 - p$.

\subsection{The method of transfer matrix}

It should be recalled that we are interested in determining the probability 
distribution of the maximum power $n$ (averaged over various types of 
intersections in the knot diagram) of the Jones-Kauffman invariant.  According
to definition  (\ref{eq:exp}), the coefficient at the term of the maximum 
power $n$ in the Jones polynomial of a randomly generated knot is 
insignificant. Let us fix a certain value  $n_0$, generate an ensemble of 
random knot diagrams, and determine the fraction of knots the polynomials of
which contain the maximum power $n=n_0$.

The traditional approach to numerical analysis of a $q$-component Potts spin 
system assumes that each possible state of the column $L_v$ spins corresponds 
to an eigenvector of the $q^{L_{v}}$-dimensional transfer matrix.  With
computers of  the P-II-300 type and a reasonable computational time, the
maximum possible  size of the transfer matrix is 100-200. This corresponds to a
spin strip  width $L_{v}=8$ for $q = 2$ and $L_v = 4$ for $q = 3$. Obviously,
this method  cannot be used for investigation of the Potts model with large $q$
values.

A method of transfer matrices applicable to the case of arbitrary $q$ was 
developed by Bl\"ote and Nightingale\cite{BN}. The basic idea of this method is 
that each eigenvector corresponds to a subdivision of the column into clusters
of "coupled" spins. By definition, the spins in each cluster are  parallel.
Therefore, the number of "colors" $q$ can be considered simply as  a parameter
taking any values (including non-integer and even complex).  A  detailed
description of this method can be also found in \cite{JC,SS}, where  this
approach was used to study the Potts model with impurities on the  bonds and to
determine the roots of a partition function for the Potts model in the complex
plane.

Below we will briefly describe the principles of determining a basis set in 
the space of coupled spins and constructing transfer matrices in this  basis.
This description follows the results reported in~\cite{SS}.

A basis set corresponds to a set of various subdivisions of the spin column 
into clusters of coupled spins.  We must take into account that the 
subdivision should admit realization in the form of a flat graph on a 
half-plane.  Now we can describe a recursive procedure for determining the 
basis set. Let a basis set be available for a column with the height $L_{v}$. 
Upon adding one more spin from below, we obtain a column with the height
$\tilde{L}_{v}$.  Then we take, for example, a subdivision  $v_{1}(L_{v})$ 
corresponding to the  first basis vector $L_v$ and begin to generate
subdivisions corresponding to  the basis vectors for the column $\tilde{L}_{v}$
by attaching the added spin to those  existing in  $v_2(L_{v})$. In each step,
we check for the possibility of  attaching this spin within the framework of
the flat graph on a half-plane.  after the attempt at attaching the added spin
to all existing clusters, we  generate a subdivision in which the added cluster
is separate.  Then we  take the next subdivision (for example, $v_2(L_{v})$)
for the $L_{v}$  column and  repeat the generation procedure.

The procedure begins with a single spin as depicted in Fig.\ref{figtm1}. This
spin is  assigned the index of unity as belonging to the first cluster. Then
another  spin is added from below to the same column. Accordingly, the strip of
two  spins may occur in one of the two states: $v_{1}(L_{v}=2)$, in which the
spins  are coupled and belong to cluster 1, or $v_{2}(L_{v}=2)$, in which the
spins are  not coupled and belong to clusters 1 and 2 (Fig.\ref{figtm1}). A
basis set for the strip of three spins is obtained by adding another spin from
below and  coupling this spin to all clusters. The clusters are enumerated by
integers  top to bottom, while vectors in the basis set are enumerated in the
order  of generation. As can be readily seen, the first basis vector
corresponds  to a state in which all spins are coupled to each other and belong
to the  same cluster, while the last basis vector corresponds to the state
where  all spins are uncoupled and the number of clusters equals to the number
of  spins in the column. The total number of vectors in the basis set for a 
strip of $m$ spins is determined as the Catalan number
$C_{m}=\frac{1}{m+1}\left({2m}\atop{m}\right)$ with the corresponding  index
number (see \cite{SS}).

Let us denote by $v_{i}(\backslash k)$ a subdivision obtained upon  separating
the $k$th spin  from the subdivision $v_{i}$ to form a separate cluster;
$v_{i}(\{k,l \} )$ will denote  a subdivision in which the clusters containing
$k$th and $l$th spins are  combined to form a common cluster. Following a
method described in \cite{SS},  let us form the matrix
$$
D_{i,j}(k)= \delta(v_{j}(\backslash k),v_{i})+
q\delta(v_{j}(\backslash k),v_{j})
$$
As can be seen, an element of the matrix $D_{i,j}(k)$ is non-zero if the $k$th 
spin is not coupled to any other spin in the subdivision  $v_{i}$. When the
same  spin is not coupled to any other in the $j$th subdivision as well, the
matrix element is $q$, otherwise it is taken equal to unity. We will also
introduce  the matrix $C(k,l)$ defined as
$$
C_{i,j}(k,l)=\delta(v_{j}(\{k,l\}),v_{i})
$$
the elements of which are equal to unity if  and only if the subdivision
$v_{i}$ is obtained from the subdivision $v_{j}$ by  combining clusters
containing the $k$th and $l$th spins.

Now let us consider a lattice structure with the first two rows such as 
depicted in Fig.\ref{figb}d (for an even index  $j+1$ in the second column) and
Fig.\ref{figb}e (for an odd index  $j+1$ in the second column). The spin
columns are  enumerated by index $j$: $1\le j \le L_{h}$, while the spin
position in a column is  determined by index  $i$: $1 \le i \le L_{v}$.

According to these constructions, the transfer matrices of the type 
$T^{even}(j+1=2k)$, which corresponds to adding an even column $j+1=2k$ after 
an odd one $j=2k-1$ (we add the columns from the right-hand side),  contain
bonds between the spins $\sigma_{i,j=2k-1}$ and $\sigma_{i+1,j=2k}$. The
transfer  matrices of the type $T^{odd}(2j+1)$, which corresponds to adding an
odd  column  $j+1=2k+1$ after an even one  $j=2k$, contain bonds between the 
spins $\sigma_{i,j=2k}$ and $\sigma_{i-1,j+1=2k+1}$.

Let the contribution to a statistical weight corresponding to the f-bond 
$b(\sigma_{i,j},\sigma_{i',j'})=1$ be expressed as 
$$
t=\exp(\beta J)
$$
where $\beta=\frac{1}{T}$. Then the a-impurity $b(\sigma_{i,j},
\sigma_{i',j'})=-1$ corresponds to the weight 
$$
t^{-1}=\exp(-\beta J)
$$
and the total  contribution of a given configuration to the statistical weight
of the  system is $t^{b(\sigma_{i,j},\sigma_{i',j'})}$. Let us define a matrix
for the horizontal bond  $(\sigma_{i,j},\sigma_{i,j+1})$ as
$$
P(i,j)= I (t^{b(\sigma_{i,j},\sigma_{i,j+1})}-1)+D(i,j)
$$
and the matrices for "sloped" bonds as 
$$
\begin{array}{l}
R^{even}(i,j)=I+C(i,j)(t^{b(\sigma_{i,j},\sigma_{i+1,j+1})}-1) \\
R^{odd}(i,j)=I+C(i,j)(t^{b(\sigma_{i,j},\sigma_{i-1,j+1})}-1)
\end{array}
$$
where $I$ is the unit matrix.

Then the transfer matrices can be expressed as follows: 
$$
\begin{array}{rcl}
T^{even}(j) & = & P(L_{v},j) R^{even}(L_{v}-1,j)P(L_{v}-1,j) \dots
R^{even}(1,j) P(1,j) \\
T^{odd}(j) & = & P(1,j) R^{odd}(2,j)P(2,j) \dots R^{odd}(L_{v},j)
P(L_{v},j)
\end{array}
$$
and the partition function is 
$$
Z(t)= u^{T}\prod \limits_{k=1}^{L_{h}/2-1}  T^{even}(L_{h})
\left( T^{odd}(2k+1) T^{even}(2k)\right) v_{max}
$$
where $v_{max}$ is a basis vector with the maximum number corresponding in our
representation to the state in which all spins belong to different clusters and
$u^{\top}= \{q^{N_{CL}(v_{1})}, q^{N_{CL}(v_{2})},..., q^{N_{CL}(v_{max})}\}$
so that $u^{\top} v_{i}=q^{N_{CL}(v_{i})}$, where $N_{CL}(v_{i})$ is the number
of various clusters in the subdivision  $v_{i}$. For  the first basis vector,
$N_{CL}(v_{1})=1$ (a single cluster) and the last vector  corresponds to
$N_{CL}(v_{max})=L_{v}$ --  $L_{v}$ different clusters.

The general algorithm of the calculation is as follows.
\begin{enumerate}
\item Generate basis set vectors for a spin strip of the required width;
\item Use this basis set to generate the matrices  $R(i,j)$, $P(i,j)$ in cases 
of f- and a-bonds ($\sigma_{i},\sigma_{j}$)
\item Generate a distribution of impurities on the lattice using a random 
number generator;
\item Generate the transfer matrices for the Potts spin model and the 
polynomial invariant;
\item Calculate a partition function for the Potts model, the polynomial 
invariant, the minimum energy, and the maximum power of the polynomial.
\item Repeat points 3-5 and perform averaging over various realizations of the 
distribution of impurities on the lattice.
\end{enumerate}

The error was determined upon averaging over ten various series of 
calculations. For each realization of the impurity distribution, the  program
simultaneously determines both the polynomial invariant and the  partition
function of the Potts model for an arbitrary (but fixed) $q$  value. This
method allows us to study correlations between the maximum  power of the
polynomial and the minimum energy within the framework of the Potts model on
the corresponding lattice.

\section{Results of calculations}

\subsection{Correlations between the maximum power of Jones polynomial
of the lattice knot and the minimum energy of the Potts model}
\label{seccor}

Let us consider dependence of the maximum power of the Jones polynomial on  the
type of the partition function for the Potts model. Formula (\ref{eqjn})
explicitly relating the Jones-Kauffman invariant to the Potts model shows 
that, if the variable $x$ were not entering into an expression for the 
degeneracy of the energy level $H(E;q=2+x+x^{-1})$, the maximum power  of the
polynomial invariant would always correspond to a term of the  partition
function with the minimum energy.

Taking into account dependence of the degree of degeneracy on the variable 
$x$, we can see that contributions to the coefficient at this power in some 
cases mutually cancel each other and this coefficient turns zero. A simple 
example is offered by a system free of impurities, in which case the  minimum
energy is  $E_{min}=-(N+1)/2$ and the Jones polynomial is identically unity
(with the maximum power being zero).

Nevertheless, there is a strong correlation between the maximum power $n$ of 
the Jones polynomial and the minimum energy $E_{min}$ of the corresponding
Potts  model. Since the minimum energy $E_{min}$ of the Potts model is always 
sign-definite and cannot be positive, we use below a positive quantity 
representing the absolute value of the minimum energy $|E_{min}|=-E_{min}$.
Figure \ref{figcorr}a shows a joint probability distribution
$P\left(\frac{n}{N},\frac{|E_{min}|}{N} \right)$ of the  normalized maximum
power $\frac{n}{N}$ and the normalized minimum energy $\frac{|E_{min}|}{N}$
obtained for a lattice with $N = 49$ and an impurity concentration of $p = 0.5$
by averaging over $N_L = 10^5$ impurity configurations.  Here and  below we use
only the normalized quantities determined on the $[0, 1]$  segment for the
energy and $[-1, 1]$ for the maximum power of the polynomial (the latter value
can be negative). This allows us to plot the curves for  various lattice
dimensions on the same figure. In Fig.\ref{figcorr}a, the probability 
distribution is described by the level curves, the spacing between which
corresponds to a probability difference of 0.001.  As can be seen,  there is a
strong correlation between the maximum power of the polynomial  invariant and
the minimum energy of the corresponding Potts model. This relationship can be
quantitatively characterized by the the coefficient of correlation.

It should be recalled that the coefficient of correlation between random 
quantities $x_{1}$ and $x_{1}$ with mathematical expectations 
$\left<x_{1}\right>$ and $\left<x_{2} \right>$ and  dispersions $\Delta
x_{1}=\left<x_{1}^{2}\right>-\left<x_{1} \right>^{2}$ and $\Delta
x_{2}=\left<x_{2}^{2}\right>- \left<x_{2} \right>^{2}$, (where $\left<\dots
\right>$ denotes averaging) is determined by the formula
$$
{\rm corr}(x_{1},x_{2})=\frac{\left<x_{1}x_{2}\right>-
\left<x_{1}\right>\left<x_{2}\right>} {\sqrt{\Delta x_{1} \Delta x_{2}}}
$$
The correlation coefficient equal to $\pm 1$ corresponds to a linear 
relationship between $x_{1}$ and $x_{2}$.

The values of the coefficient of correlation between the maximum power
$n_{max}$  of the polynomial invariant  and the minimum energy $|E_{min}|$ for
the impurity  concentration $p = 0.5$ on lattices with various the linear
dimensions  $L = 3, 5, 7, 9$, and 11 are presented in Table 1. 
These values were obtained by  averaging over $N_{L} = 10^{5}$ impurity
realizations for $L = 3\div 7$, $N_L = 2 \times 10^4 $ for $L = 9$, and $N_L =
10^3$ for $L = 11$. As is seen, the correlation increases with the lattice
size.

\medskip

{\small TABLE 1. Mean value of the correlation coefficient ${\rm corr}\left(
n_{max},|E_{min}| \right)$  and the corresponding statistical error
for several lattice sizes.}

\begin{center}
\begin{tabular}{|c|c|c|}
\hline
$L$ &
${\rm corr} \left( n,|E_{min}| \right) $ &
$\Delta {\rm corr} \left( n,|E_{min}| \right) $ \\
\hline
3 & 0.4871 &0.0021\\
\hline
5 & 0.6435 &0.0022\\
\hline
7 & 0.7205 &0.0007\\
\hline
9 & 0.7692 &0.0013\\  
\hline
11& 0.7767 &0.0129\\
\hline
\end{tabular}
\end{center}

Figure \ref{figcorr}b shows the approximation of these data by a power function
${\rm corr}  \left( \frac{n}{N},\frac{|E_{min}|}{N}\right)
=1.04(4)-1.126(14)L^{-0.65(7)}$. Naturally, the correlation coefficient cannot
be greater than unity. The  last expression is the result of ignoring the
higher terms of expansion in  powers of  $\frac{1}{L}$. The approximation shows
that there is a relationship between  the maximum power of the polynomial
invariant and the minimum energy of  the Potts model, the degree of correlation
increasing with the lattice  size. In what follows, the results for the maximum
power of the polynomial  invariant will be accompanied by data for the minimum
energy.

The presence of a correlation between the maximum power $n$ of the Kauffman 
polynomial invariant and the minimum energy  $E_{min}$ of the corresponding
Potts  model for each particular impurity realization on the lattice is of 
interest from theoretical standpoint and can be used in the numerical 
experiments. At present, the limiting lattice size $N_L$ used in the method of 
transfer matrix does not exceed $N_L = 11$, which is determined by a large
volume of necessary computations. At the same time, the minimum energy for  the
Potts model can be calculated within the framework of the standard  Monte-Carlo
algorithm, which poses a much smaller requirements to the computational
facilities and, hence, can be readily applied to the  lattices of significantly
greater size.

\subsection{The probability distribution of the maximum power of the polynomial
invariant and the minimum energy of the corresponding Potts model}

Here we present the results of determining the probability distribution  $P(n)$
of the maximum polynomial power $n$. Figure~\ref{figgik7}a shows the $P(n)$
curves for  $L = 7$ and the impurity concentrations $p = 0.1$ (depicted by
crosses), 0.2  (squares), and 0.5 (circles). The data were obtained by
averaging over  $N_L = 10^5$ impurity realizations.  The statistical errors are
smaller than the  size of symbols.

As can be seen, the $P\left(\frac{n}{N}\right)$ curve for small impurity
concentrations is  nonmonotonic. As the $p$ value increases, the probability
distribution  function becomes monotonic and approaches in shape to the Gauss
function.  Figure \ref{figgik7}b shows the corresponding probability
distribution of the  normalized minimum energy $\frac{|E_{min}|}{N}$ obtained
for the same lattice size ($L = 7$)  and impurity concentrations (denoted by
the same symbols). This function  appears as more monotonic and approaches the
Gauss function already at  small impurity concentrations ($p\simeq 0.2$).

The shapes of the probability distribution observed for a fixed impurity 
concentration ($p = 0.5$) and various lattice dimensions are shown in
\ref{figmomk50}a  (for the maximum polynomial power) and Fig.\ref{figmomk50}b
(for the minimum energy).  Data for the lattice size $L = 3$ (squares), $5
(squares)$, and 7 (circles),  and 9 (triangles) were averaged over $N_L = 10^5$
($L = 3, 5,$ and 7) and data  for $L = 9$, over $N_L = 5 \times 10^3$ impurity
realizations. The probability was normalized to  unity:  $\sum \limits_{n}
P(n)=1$ For this reason, an increase in the lattice size is accompanied by
growing  number of the values that can be adopted by the normalized maximum
power of  $P(\frac{n}{N})$, while the value of the probability distribution
decreases  approximately as $\frac{1}{N}$. As is seen, the probability
distribution $P\left(\frac{n}{N}\right)$ for  lattices of smaller size ($L = 3$
and 5) is nonmonotonic; as the $L$ value  increases, the distribution becomes a
smooth Gaussian-like function.

Thus, we may conclude that the probability distribution
$P\left(\frac{n}{N}\right)$ of the maximum  power of the polynomial invariant
for small-size lattices is nonmonotonic  as a result of the boundary effects
even for a considerable impurity  concentration. As the lattice size increases,
the nonmonotonic character  disappears and the probability distribution becomes
a smooth function.  Apparently, we may ascertain that, whatever small is the
impurity concentration $p$, there is a lattice size $N$ ($N \gg \frac{1}{p}$)
such that the  corresponding probability distribution is smooth and
Gaussian-like. Thus,  we may suggest that the probability distribution $P(n)$
on the lattices of  large size is determined by the Gauss function, the main
parameters of  which are the mathematical expectation (mean value) of the
maximum polynomial power  and the dispersion. Plots of the mean value of the
maximum  polynomial power $\left<\frac{n}{N}\right>$ and the corresponding
dispersion  $W^{2}_{knot}$ as functions of the impurity concentration $p$ are
presented in Figs. \ref{figgik357}a  and Fig.\ref{figgik357},  respectively,
for the lattices size of $L = 3$ (crosses), 5 (squares), and 7  (circles). The
data for each point were obtained by averaging over $N_{L}  = 10^{5}$ impurity
realizations for $L = 3$ and 5 and $N_L = 5 \times 10^4$ for $L = 7$.

As expected, both the mean value and the dispersion of the maximum  polynomial
power turn zero for $p = 0$ and 1 (trivial knot) and reach maximum  at $p
\simeq 0.5$. Note that these functions are not symmetric with respect to the 
transformation $p \to 1 - p$, since the Jones polynomial invariant of a mirror 
knot (with all overcrossing changed for undercrossing and vice versa) is 
obtained by changing variables $x \to x^{-1}$ , whereby the maximum polynomial 
power of the mirror knot corresponds to the minimum power of the original 
polynomial, taken with the minus sign. However, this asymmetry disappears  with
increasing lattice size $N$ as a result of increase in the amount of 
impurities and in the number of possible impurity realizations employed in  the
averaging.  For the comparison, Fig.\ref{figgik357}b shows the function 
$\frac{1}{2}p(1-p)$ representing the dispersion of the distribution function in
the  hypothetical case when the maximum polynomial power $n$ is a linear
function  of the number of impurities $M$. Thus, a difference between
$W^{2}_{knot}$ and $\frac{1}{2} p(1-p)$ characterizes dispersion of the
distribution of the maximum polynomial  power at a fixed number of impurities
$M$ (it should be recalled that the  impurity occupies each lattice site with a
probability $p$ and the total  number of impurities $M$ fluctuates).

Figures \ref{figgip357}a and \ref{figgip357}b (with the parameters and
notations analogous to those  in Fig.\ref{figgik357}a and Fig.\ref{figgik357}b)
shows data for the absolute value of the mean minimum energy 
$\left<\frac{|E_{min}|}{N}\right>$  and the corresponding dispersion 
$W^2_{Potts}$. The asymmetry of the mean  minimum energy plot is related to the
fact that the number of f-bonds is  greater than that of the a-bonds by one at
$p = 0$ and is smaller by one, at  $p = 1$.  The probability distribution of
the minimum energy is treated in more detail in Subsection  \ref{razpot} of the
Appendix.  We have studied  dependence of the mean normalized maximum
polynomial power on the lattice  size $L$ for $p = 0.5$ (Fig.\ref{figmomk50}a).
The results were averaged over $N_L = 10^5$ impurity realizations for $L = 3,
5$, and 7 and over $N_L = 1.5 \times 10^4$ realizations for $L = 9$. As can be
seen, the  $\left<\frac{n}{N} \right>$ value tends to a certain  limit with
increasing $L\to \infty$ . We approximated the results by a power function to
obtain
\be
\label{eq:fit_n}
\left<\frac{n}{N} \right> \simeq 0.334(8)-0.41(2)L^{-0.48(5)}
\ee
Analogous data for the normalized minimum energy in the Potts model are 
depicted in Fig.\ref{figmomk50}b. The mean absolute value of the minimum
energy  decreases with increasing lattice size and can be approximated by a
power  function of the type
$$
\left<\frac{|E_{min}|}{N}\right> \simeq 0.4185(7)+ 0.119(3) L^{-1.11(4)}
$$
Some features of the minimum energy distribution in the Potts model with 
random ferro- and antiferromagnetic bonds are treated in the Appendix for 
$q\ge 4$, in which case analytical expressions can be obtained for small ($p
\simeq 0$)  and large  ($p \simeq 1$) impurity concentrations.

\section{Conclusions}

The results of this investigation can be formulated as follows.

1. An analysis of the relationship between the partition function of the  Potts
model and the Jones-Kauffman polynomial invariant, in combination  with the
results of numerical calculations for the Potts model, allowed us  to study the
probability distribution $P(\frac{n}{N};p,N)$ of the maximum  polynomial power
$n$ for various concentrations (probabilities) $p$ of an  "impurity"
representing intersections of the  $\epsilon=-1$ type in the sites of a square
lattice with $N = L \times L$ for $L = 3-11$.

2. For the lattices of small size with small impurity concentrations $p$, the 
probability distribution  $P(\frac{n}{N};p,N)$ of the normalized maximum 
polynomial
invariant  power $n$ is not smooth, showing an alternation of more and less
probable states. This behavior indicates that the knots of certain topological
types  are difficult to realize for $p  \ll 1$ on a square lattice as a result
of  geometric limitations.

3. As the impurity concentration (probability) $p$ increases, the probability 
distribution $P(\frac{n}{N};p,N)$ of the normalized maximum polynomial 
invariant power
$n$  becomes smooth, with the shape approaching that of the Gauss function. A 
typical value of the "knot complexity"  $\eta$ for $p = 0.5$ and sufficiently
large  lattices ($N \gg 1$) can be obtained by extrapolating the expression
(\ref{eq:fit_n}):
$$
\eta=\lim_{N\to\infty} \left<\frac{n}{N}\right>\approx 0.334
$$

4. There is a correlation between the maximum power of the polynomial 
invariant of a knot and the minimum energy of the corresponding Potts 
lattice model for $q  \ge 4$.

5. An analytical expression was obtained for the probability distribution 
function $P(\frac{E_{min}}{N};p,N)$ at relatively small $p\sim \frac{1}{N}$
impurity  concentrations.

We have studied the knots with a square lattice diagram. However, all 
calculations can be generalized to the case of rectangular lattices. In 
connection with this, it would be of interest to check whether the 
distribution of knots over the topological types only depends on the number  of
intersections on the lattice diagram or it depends on the diagram shape  as
well. We may expect that the probability distribution function for a  strongly
elongated rectangular diagram would differ from the distribution for a square
diagram with the same number of intersections.

The method developed in this study is applicable, in principle, to the 
investigation of knots with arbitrary diagrams (including the case of knots 
with the diagrams not tightly fit to a rectangular lattice), provided that  a
configuration of the Potts system corresponding to this diagram is known.

We believe that the proposed combination of analytical and numerical  methods
for the investigation of topological problems using the models of  statistical
physics offers both a promising means of solving such topological problems and
a new approach to the standard methods of  investigation of disordered
systems.  This can be illustrated by the  following fact. One of the main
concepts in the statistical physics is the  principle of additivity of the free
energy of a system, that is,  proportionality of the free energy to the system
volume $N$.  By  interpreting the free energy as a topological characteristic
of the  "complexity" of a knot, we may conclude that the complexity of a
typical  knot increases linearly with the system volume $N$. This property is,
in  turn, well known in the topology (outside the context of statistical
physics), being a fundamental manifestation of the non-Abelian 
(non-commutative) character of the phase space of knots.

\bigskip

\centerline{\bf Acknowledgments}
\bigskip

This study is a logical development of the ideas formulated by one of the 
authors in collaboration with A.Yu. Grosberg in 1992-1993 \cite{Gr_Ne}. The
authors  are grateful to A.Yu. Grosberg for fruitful comments and to J.L. 
Jacobsen for helpful discussions concerning transfer--matrix approach. The
study was partly supported by the russian Foundation for Basic  Research,
project no.  00-125-99302. One of the authors (O.A.V.) is  grateful to the L.D.
Landau Grant Committee (Forschungzentrum, KFA Julich,  Germany) for support.

\begin{appendix}

\section*{Minimum energy distribution in the Potts model with random ferro- and 
antiferromagnetic bonds}

\subsection{Independence of the minimal energy upon $q$ for $q \ge 4$}
\label{secq}

The minimum energy of a spin system is independent of $q$ for $q  \ge 4$. In 
this study of the topological invariants of random lattice knots, we are 
interested in determining behavior of the system for $x \to \infty$(see Eq.
(\ref{eq:exp})).  With an allowance for the relationship $q=2+x+x^{-1}$, this
implies $q \to \infty $ for the corresponding Potts model. Some features of the 
dependence of the free energy of the Potts spin system on the number of  states
$q$ can be established based on simple considerations.  In cases when  the
lattice contains no impurities and the a-bonds are arranged as depicted  in
Figs.\ref{fig1}b and \ref{fig1}c, the minimum energy is independent of $q$ for
$q \ge 2$. Under these conditions, two spin values $\sigma=\{1,2\}$ are
sufficient to "create"  a configuration corresponding to the minimum energy. In
the presence of  impurities, the minimum energy depends on $q$ for $q \ge 2$.
This is illustrated in Fig.\ref{figb}a for a $5 \times 5$ lattice with a single
impurity on the bond between  $\sigma_{1,1}$ and $\sigma_{1,2}$  spins. For the
spin configuration energy to reach a minimum  value of $E_{min}=-12$ for the
given distribution of bonds, it is necessary that all spin variables in the
lattice sites connected by f-bonds $J=1$ (in Fig.\ref{figb}, these sites are
indicated by open and filled circles) would  acquire the same values, while the
spin variables in the sites $\sigma_{1,1}$ and $\sigma_{1,2}$ connected by
a-bonds  $J=-1$ (in Fig.\ref{figb}a, these sites are indicated by filled and
hatched circles) would acquire different values  (e.g., $\sigma_{1,1}=2,\;
\sigma_{1,2}=3$). However, it is impossible to assign the values  of spin
variables in a system with $q = 2$ so that the spins in clusters
$\sigma_{1,1},\;\sigma_{1,2}$ (as well as in the adjacent cluster) indicated by
unlike symbols (black versus open or hatched) were different.  For this reason,
a  minimum energy of the spin state for $q = 2$ is $E_{min}=-11$ (instead of
$-12$).

As was demonstrated, a minimum energy of the spin configuration in the  Potts
model corresponding to a given distribution of a-bonds may depend on  the
number of spin states $q$. This fact can be represented as follows: to  reach
the state with minimum energy, it is necessary that the spin  variables in the
lattice sites connected by f-bonds would acquire for the  most part the same
values (so as to form clusters), while the spin  variables in the sites
connected by a-bonds would be possibly different.   Thus, a given distribution
of the a-bonds corresponds to a subdivision of  the lattice into independent
clusters of spins. Spins belonging to the same  cluster are connected
predominantly by the f-bonds, while spins of  different clusters are connected
by a-bonds.

The energy reaches minimum if the adjacent clusters in a given subdivision 
possess different values of the spin variable. We may bring each value of  the
spin variable into correspondence with a certain color. Then a minimum  energy
corresponds to the lattice subdivision into clusters painted so  that all spins
in one cluster are of the same color, whereas adjacent  clusters have different
colors. Figure~\ref{figb}a shows an example of the configuration which cannot
be painted in this way using two colors, while  three colors allow reaching the
goal.  In mathematics, there is a theorem  concerning the task of "painting
maps", according to which any  configuration on a surface possessing a a
topology of a sphere can be painted using four (or more) colors. In other
words, any subdivision  of a lattice into clusters can be painted using four
(or more) colors so  that the adjacent clusters would possess different colors.
If each color corresponds to a certain value of the spin variable
$q=\{1,2,3,4,...\}$, we may  assign the $q$ values (for $q \ge 4$) so that
spins in the adjacent clusters would  possess different values (colors). There
is no impurity configuration (and the corresponding lattice subdivision into
clusters) such that four values  of the spin variable would be insufficient to
reach the state of minimum  energy.

The above considerations allow us to formulate the following statement: For  an
arbitrary configuration of a-impurities on a Potts model lattice, a  minimum
energy of the spin system is independent of the number of spin  states $q$ for
$q \ge 4$.

As is known, the Potts model exhibits a first-order phase transition at  $q 
\ge 4$. Note also that the parameter in the Kauffman invariant becomes real 
just for $q \ge 4$.

\subsection{Distribution of the minimum energy at small ($p \simeq 0$) and
large ($p \simeq 1$) impurity concentrations}
\label{razpot}

Consider the Potts model on an $L_{v} \times L_{h}$ lattice with a total number
of  lattice sites $N$ and $q \ge 4$. Let $M$ be a fixed number of impurities of
the $\epsilon_{k}$  types. In the absence of impurities ($M = 0$), the number
of f-bonds is $N_{+}=\frac{N+1}{2}$ and  the number of a-bonds is 
$N_{-}=\frac{N-1}{2}$. The arrangement of a-bonds is illustrated in
Fig.\ref{figb}a. The minimum energy in the absence of impurities is
$E_{min}=-N_{+}$.

\medskip

{\small TABLE 2. The probability distribution of energy $P(E_{min};M,N)$ 
for a fixed number of impurities $M=0,1,2,N-2,N-1,N$ on a lattice with 
$N$sites ($C=\sqrt{N}$).}   

\begin{center}
\begin{tabular}{|c||c||c||c||}
\hline
I & II & III & IV 
\\ \hline \hline
$M$& Link type& $E_{min}$ & $P(E_{min};M)$ 
\\ \hline \hline
0 &  &$-\frac{N+1}{2}$ & $1$ 
\\ \hline
1 &$+$ & $-\frac{N+1}{2}+1$ & $ \frac{N+1}{2N}$  
\\ \hline
1 &$-$  &$ -\frac{N+1}{2}$ & $ \frac{N-1}{2N}$  
\\ \hline
2 &$++$ &$-\frac{N+1}{2}+2$ & $\frac{(N+1)(N-1)}{4N(N-1)}$ 
\\ \hline
2 & $+-$&$-\frac{N+1}{2}+1$ &
$\frac{(N+1)(N-1)-16(C-1)}{2N(N-1)}$ 
\\ \hline
2 & $+-$\&$--$ & $-\frac{N+1}{2}$ &
$\frac{(N-1)(N-3)+16(C-1)}{4N(N-1)}$\\
\hline 2&$--$ & $-\frac{N+1}{2}-1$ & $\frac{4(C-2)}{N(N-1)}$  \\
\hline 2 &
$--$ & $-\frac{N+1}{2}-2$ & $\frac{4}{N(N-1)}$  
\\ \hline
N-2 & $++$ & $-\frac{N-1}{2}+2$ & $\frac{(N-1)(N-3)}{4N(N-1)}$ \\ 
\hline
N-2 & $+-$ & $-\frac{N-1}{2}+1$ &  
$\frac{(N-1)(N-3) -16(C-1)}{2N(N-1)}$ \\ \hline
N-2 & $+-$\&$--$ & $-\frac{N-1}{2}$ &
$\frac{(N+1)(N-1) +16(C-1)-4}{4N(N-1)}$ \\ \hline
N-2 & $+-$\&$--$ & $-\frac{N-1}{2}-1$ &
$\frac{2(N-1)+2(C-1))}{N(N-1)}$\\ 
\hline
N-2 & $--$ & $-\frac{N-1}{2}-2$ & $\frac{2}{N(N-1)}$ \\ \hline
N-1 & $+$ & $-\frac{N-1}{2}+1$ & $\frac{N-1}{2N}$ \\ \hline
N-1 & $-$ & $-\frac{N-1}{2}$ & $\frac{N-3}{2N}$ \\ \hline
N-1 & $-$ & $-\frac{N-1}{2}-1$ & $\frac{2}{N}$ \\ \hline
N &  & $-\frac{N-1}{2}$ & 1 \\   
\hline
\end{tabular}
\end{center}

We may calculate the probability distribution $P(E_{min};M,N)$ of the minimum 
energy for $M = 1$ and 2 by trials of the possible variants. In the case of a 
single impurity ($M = 1$), there is a probability $N_{+}/N=(N+1)/2N$ for this
impurity to fall on the f-bond, after which the minimum energy  increases by
unity:   $P(E=-N_{+}+1)=(N+1)/2N$ There is also the probability
$N_{-}/N=(N-1)/2N$ for the impurity to fall on the a-bond, after which the
minimum energy  remains unchanged: $P(E=-(N+1)/2)=(N-1)/N$ Similar
considerations can be conducted for $M = 2$. Let $C = \sqrt{N}$  denote the 
number of spin clusters with the same $q$ on a lattice without impurities.  The
results of this analysis are summarized in Table 2, where each line 
indicates (I) the number of impurities $M$, (II) the coupling constant of a 
bond on which the impurity falls (with the signs "$+$" and "$-$" corresponding 
to $b_{i,j}=1$and $-1$, respectively), (III) the minimum energy $E_{min}$
and (IV) the probability $P(E_{min};M)$ of obtaining a given minimum
energy value.

With $p$ denoting the probability that a bond contains the impurity  $J'=-1$,
the probability to find $M$ impurities is $P(M;p)=\frac{N!}{M!(N-M)!}
p^{M}(1-p)^{N-M}$. Thus, a mean value of the minimum energy $E$ in the Potts
model is 
\be \label{eqbin} \left< E_{min} \right>=\sum \limits_{M=0}^{N}
\frac{N!}{M!(N-M)!} p^{M}(1-p)^{N-M} P(E;M,N) 
\ee 
Calculating the first three terms for this sum using the data in
Table 2,  we obtain an approximate formula for the mean minimum energy
\be \label{eqpf}
\begin{array}{ll}
\left< E_{min} \right> \simeq (1-p)^{N-2} & 
\Big((1-p)^{2}P(E;M=0)+N p
(1-p)P(E;M=1) + \\ & p^{2} P(E;M=2) \Big) + o(p^{2})
\end{array}
\ee
This formula is applicable if the probability for three impurities to  appear
on the lattice is small, that is, if $\frac{1}{6}N(N-1)(N-2)(1-p)^{N-3}p^{3}
\ll 1$ Upon calculating the probability of the impurity concentration for $M =
1$  and 2 and using expression (\ref{eqbin}), we obtain a formula for
$\left<E_{min}\right>$ at small $p$.  For example, in the case of $M = 1$ this
yields
$$
\left< E_{min}(M=1) \right> = \left( -\frac{N+1}{2} \right)
\left(\frac{N-1}{2N} \right) + \left(- \frac{N+1}{2}+1\right)
\left(\frac{N+1}{2N} \right)=-\frac{N+1}{2}+\frac{N+1}{2N}
$$
By the same token, for $M = 2$ we obtain
$$
\left<E_{min}(M=2)\right>=-\frac{N+1}{2}+\frac{N+1}{N}-\frac{12C-8}{N(N-1)}
$$
Using these expressions and taking into account that $C = \sqrt{N}$ , we obtain
an  expression for the mean normalized minimum energy  $<e_{min}>=\left<
\frac{E_{min}}{N}\right>$ as a  function of the impurity concentration $p$ (in
the case of $p \ll 1$)
\be
\label{eqp0}
\begin{array}{lll}
<e_{min}(p)> & = & \frac{1}{N} \bigg[ (1-p)^{N}P(E;M=0)+N
p(1-p)^{N-1}P(E;M=1)+ \medskip \\
& & \disp \frac{1}{2}N(N-1)p^{2}(1-p)^{N-2}P(n;M=2)+o(p^{2})\bigg]=
\medskip \\
& & \disp -\frac{1}{2}\left( 1- \frac{1}{N} \right)+p   \frac{1}{2}   
\left(1-\frac{3}{N} \right)-
p^{2} \frac{6\sqrt{N}-4}{N}+o(p^{2})
\end{array}
\ee
Analogous calculations for the concentrations $1-p \ll 1$ yield
$$
\begin{array}{l}
\left<E_{min}(M=N-1)\right> =- \frac{N-1}{2} + \frac{N-5}{2N} \\   
\left<E_{min}(M=N-2) \right>=- \frac{N-1}{2} +
\frac{N-5}{N}-\frac{10C-14}{N(N-1)}
\end{array}
$$
and
\be
\label{eqp1}
\begin{array}{lll}
<e_{min}(p)> & = & \frac{1}{N} \bigg[ p^{N}P(E;M=N)+N
(1-p)p^{N-1}P(E;M=N-1)+ \medskip \\
& & \disp \frac{1}{2}N(N-1)(1-p)^{2}p^{N-2}P(n;M=N-2)+o(p^{2})\bigg]=
\medskip \\
& & \disp - \frac{1}{2} \left(1- \frac{1}{N}\right) +
(1-p)\frac{1}{2} \left(1-\frac{5}{N} \right)-
(1-p)^{2} \frac{5\sqrt{N}-7}{N}+o((1-p)^{2})
\end{array}
\ee

The results of calculations using the formulas presented in Table 2
and the  results of numerical analysis for $q = 4$, $L_v = 3$, and $L_h = 6$
(corresponding  to a knot with a $5 times 5$ square lattice diagram) are
plotted in  Fig.\ref{figgism5}a in  coordinates of the minimum energy modulus
versus probability. Figure \ref{figgism5}a shows the results for a fixed number
of impurities $M = 1$ (squares) and 2  (circles), while Fig.\ref{figgism5}b
presents data for the fixed impurity  con-centrations $p = 0.005$ (crosses),
0.01 (squares), and 0.015 (circles).  The results of calculations using the
formulas from Table 2 and  relationship (\ref{eqpf}) are depicted by
lines and the values obtained by the  Monte Carlo method are represented by
symbols. The data were averaged over  $N_L = 10^5$ realizations; the
statistical error for most of the points is  smaller than the size of symbols.
As can be seen, the numerical data  remarkably fit to the analytical curves. In
\ref{figgism5}b, the analytical  calculations are illustrated only in the
range $|E_{min}|=11 \div 15$, where the  results could be obtained by expansion
into series with terms on the order  of $p^2$ . Thus, we have numerically
verified the results of analytical  calculations performed in this section.

Let us make some remarks concerning the form of the normalized minimum  energy 
$<e_{min}(p;N)>=\left<\frac{|E_{min}(p;N)|}{N}\right>$ for  $N \to \infty$. As
can be seen in Fig.\ref{figgik357}b, the plots of 
$<e_{min}(p;N)>=\left<\frac{|E_{min}(p;N)|}{N}\right>$ for the  lattices with
$N = 25$ and 49 appear as asymmetric troughs. As $N \to \infty$, the  values 
of this function at $p = 0$ and $1$
$$
<e_{min}(p=0;N)>=0.5\left(1+\frac{1}{N}\right)
$$
and
$$
<e_{min}(p=1;N)>=0.5 \left(1-\frac{1}{N}\right)
$$
will be equal and the profiles will be symmetric with respect to the 
transformation $p \to 1 - p$. The bottom will occur on a level of $<e(p=0.5;N
\to \infty)>=0.415(7)$ (see Fig.\ref{figmomk50}b). As can be seen from
formulas  ~(\ref{eqp0})  and ~(\ref{eqp1}), the second  derivative of this
function with respect to $p$ at the points $p = 0$ and 1  turns zero for $N \to
$. Probably, all derivatives of the higher orders will  turn zero as well. In
this case, the derivative exhibits a break at the  points corre-sponding to the
trough "corners."

\end{appendix}

\newpage

\centerline{\bf Figure Captions}
\bigskip

{\small 

\begin{fig}
{\bf a)} A knot diagram on the $N = 3 \times3$ lattice and {\bf b)} the diagram
splitting. Open circles indicate the spin positions in the Potts model;  dashed
lines show the graphs on the Potts lattice.
\label{figpot}
\end{fig}
\medskip

\begin{fig}
The Reidemeister moves I, II, and II. 
\label{fig:radem}
\end{fig}
\medskip

\begin{fig}
{\bf  a)} A knot on $N = 5 \times 5$ lattice; {\bf b)} Potts spin
configuration corresponding to this lattice, and {\bf c)} the same spin
configuration reduced to a rectangular lattice $L_{ h} \times L_{v} = 3
\times6$.
\label{fig1}
\end{fig}
\medskip

\begin{fig}
A schematic diagram illustrating the step-by-step generation of subdivisions
into clusters for a column of spins corresponding to the knot state vectors.
\label{figtm1}
\end{fig}
\medskip

\begin{fig}
{\bf a)} An example of the configuration of bonds in which a minimum energy of
the Potts model is not reached for $q = 2$; {\bf b), c)} the arrangement of f-
and a-bonds on the Potts lattice with $N = 3 \times 3$ and the impurity
concentrations $p = 0$ and $p= 1$, respectively; (c, d) the arrangement of
bonds between col- umns with even and odd numbers, respectively.
\label{figb}
\end{fig}
\medskip

\begin{fig}
{\bf a)} A joint probability distribution $P \left(\frac{n}{N},
\frac{E_{min}}{N} \right )$ of the  normalized maximum polynomial power
$\frac{n}{N}$ and the normalized minimum energy  $\frac{E_{min}}{N}$ described
by the level curves with a step of $0.001$ for a lattice  with $N = 49$ and an
impurity concentration of $p = 0.5$; {\bf b)} The coefficient  of correlation
between the maximum power of the polynomial invariant and  the minimum energy
as a function of the lattice size $L$; dashed curve shows  the approximation of
these data by a power function of $L$.
\label{figcorr}
\end{fig}
\medskip

\begin{fig}
The distributions of {\bf a)} probability $P(\frac{n}{N};p)$ of the normalized 
maximum polynomial power $\frac{n}{N}$  and {\bf b)} probability $P
\left(\frac{|E_{min}|}{N};p \right)$ of the  normalized minimum energy modulus
$E_{min}$ for a $7 \times 7$ lattice and various  impurity concentrations $p =
0.1, 0.2$ and $0.5$.
\label{figgik7}
\end{fig}
\medskip

\begin{fig}
The probability of {\bf a)} the normalized maximum polynomial power 
$\frac{n}{N}$  and {\bf b)} the normalized minimum energy modulus
$\frac{|E_{min}|}{N}$  for the square  lattices with $N = 9, 25, 49$, and $81$
and an impurity concentration of  $p = 0.5$.
\label{gipl}
\end{fig}
\medskip

\begin{fig}
Plots of {\bf a)} the mean normalized maximum polynomial power 
$\left<\frac{n}{N}\right>$ and  {\bf b)} the dispersion $W_{knot}^{2}(p)$ of
the $\frac{n}{N}$ value distribution versus the impurity  concentration $p$ in
the range from 0 to 10.
\label{figgik357}
\end{fig}
\medskip

\begin{fig}
Plots of {\bf a)} the mean normalized minimum energy modulus 
$\left<\frac{|E_{min}|}{N} \right>$ and {\bf b)} the dispersion
$W_{Potts}^{2}(p)$ of the $\frac{|E_{min}|}{N}$   value distribution versus
the  impurity concentration $p$ in the range from 0 to 10.
\label{figgip357}
\end{fig}
\medskip

\begin{fig}
Plots of {\bf a)} the mean normalized maximum power $\left< \frac{n}{N}
\right>$ of the  polynomial invariant and {\bf b)} the normalized mean minimum
energy modulus  $ \left< \frac{|E_{min}|}{N} \right>$ for the square lattices
with $L = \sqrt{N}=3, 5, 7$, and $9$ and an impurity  concentration of $p =
0.5$; dashed curve in {\bf b)} shows the results of  approximation by the power
function $0.4185(7)+0.119(3)L^{-1.11(4)}$.
\label{figmomk50}
\end{fig}
\medskip

\begin{fig}
A comparison of the results obtained by analytical (anal.)  and numerical
(num.) methods for the minimum energy probability  distributions on a lattice
with $N = 5\times 5$: {\bf a)} $P(|E_{ min}|; M)$  with $M = 1$ and 2; {\bf b)}
$P(|E_{min}|;p)$  with $p = 0.005, 0.01$, and 0.015.
\label{figgism5}
\end{fig}

}

\newpage

{\topmargin =1.5in

\begin{figure}[p]
\begin{center}
\begin{picture}(0,-00)
\setlength{\unitlength}{1.000pt}
\put(85,-15){\Large\bf a)}
\put(285,-15){\Large\bf b)}
\end{picture}
\epsfxsize=130mm
\epsfysize=60mm
\epsfbox{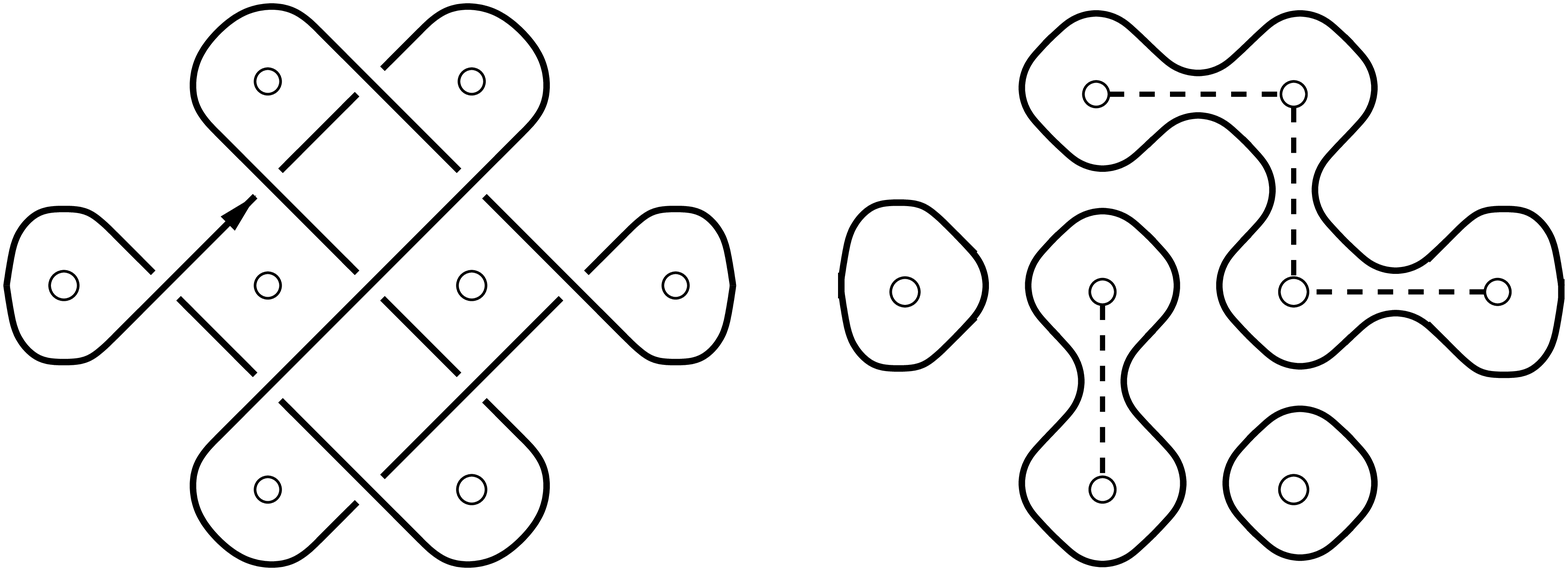}
\end{center}
\vspace{8mm}
\caption{ }
\end{figure}

\clearpage 

\begin{figure}[p]
\begin{center}
\begin{picture}(0,-00)
\setlength{\unitlength}{1.000pt}
\put(-10,+159){\large\bf I}
\put(-10,+90){\large\bf II}
\put(-10,+30){\large\bf III}
\end{picture}
\epsfxsize=90mm
\epsfysize=64mm
\epsfbox{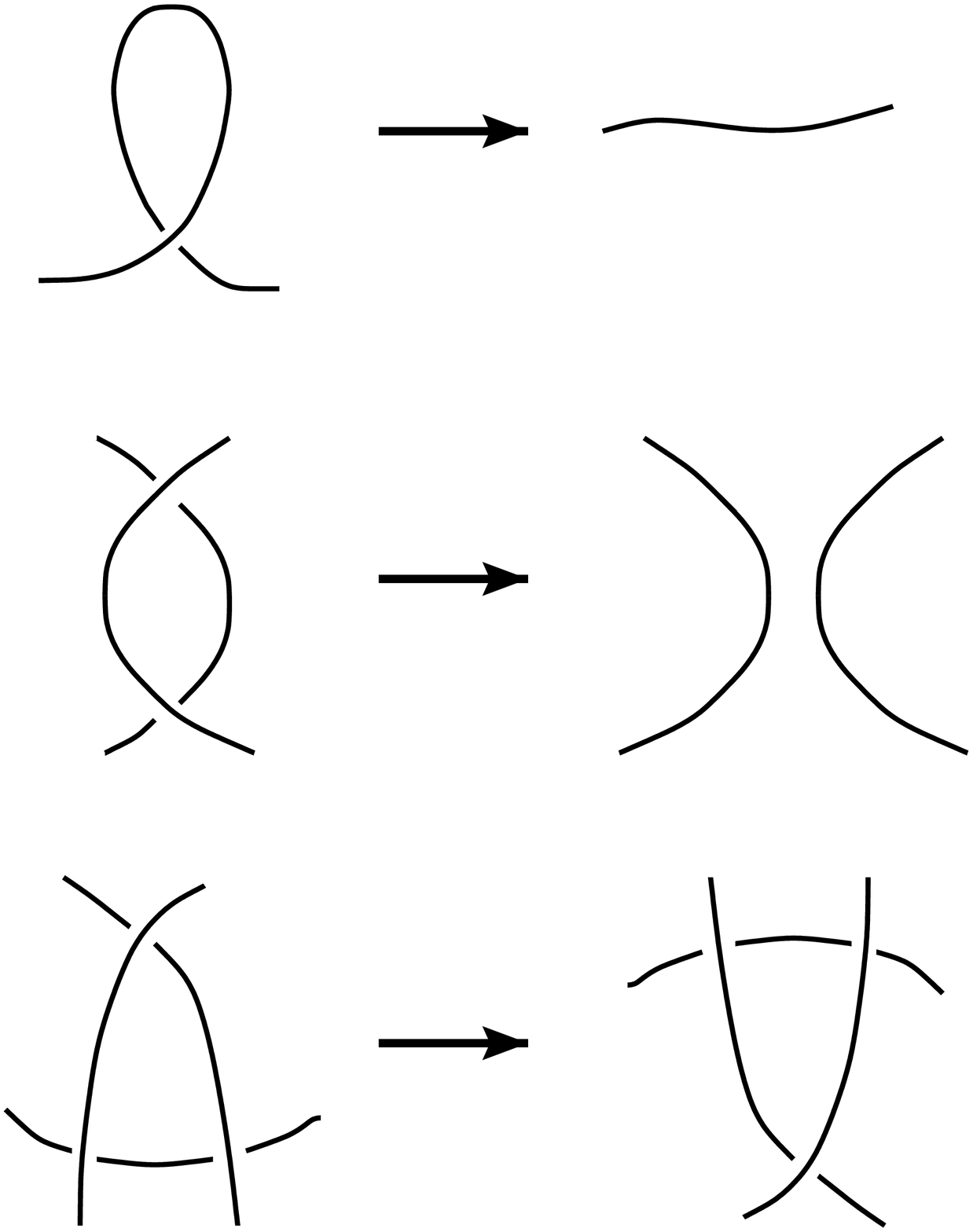}
\end{center}
\vspace{5mm}
\caption{ }
\label{radem}
\end{figure}

\clearpage

\begin{figure}[p]
\begin{picture}(400,300)
\setlength{\unitlength}{1.000pt}
\put(50,120){
\epsfxsize=50mm
\epsfysize=50mm
\epsfbox{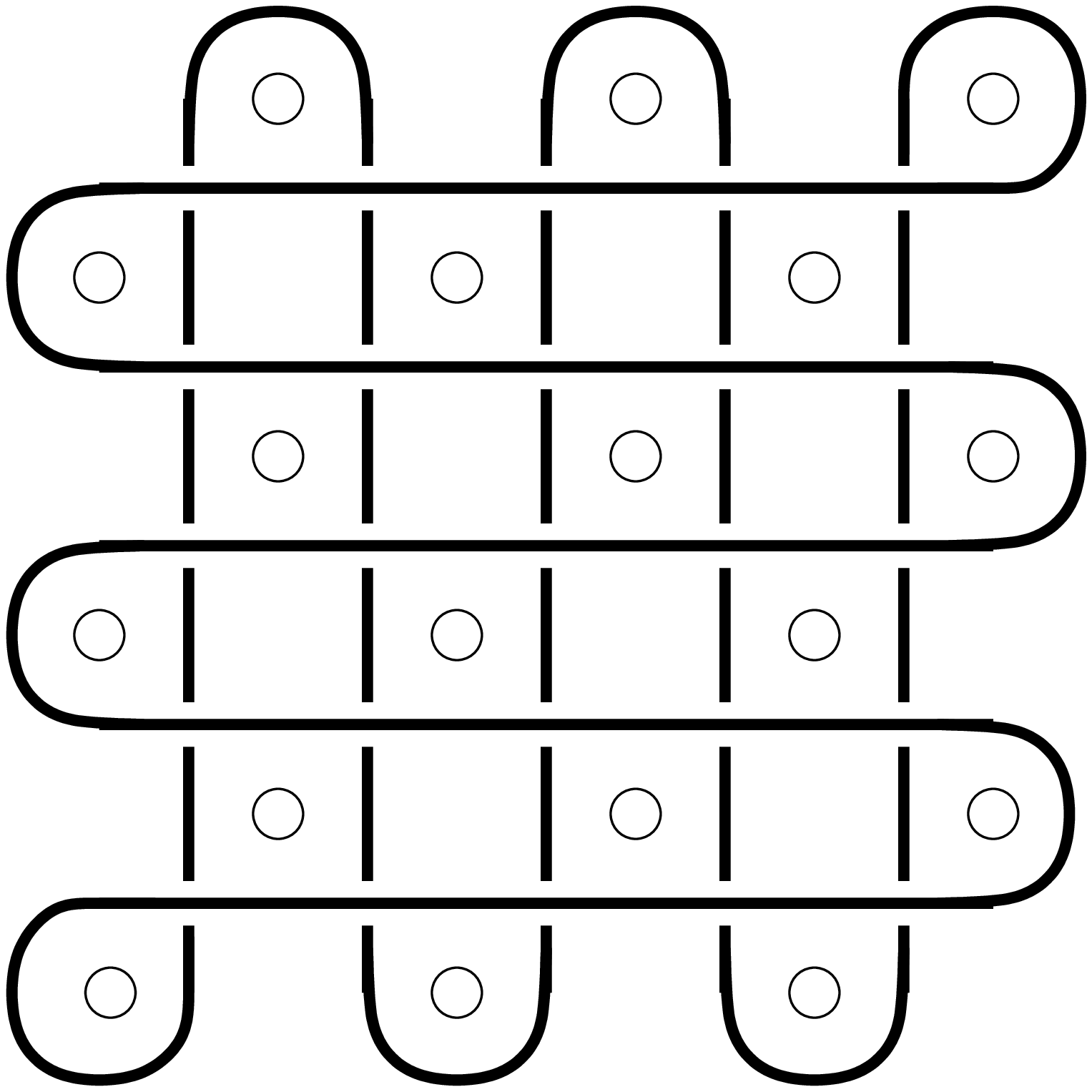}
\put(60,0){
\epsfxsize=50mm
\epsfysize=50mm
\epsfbox{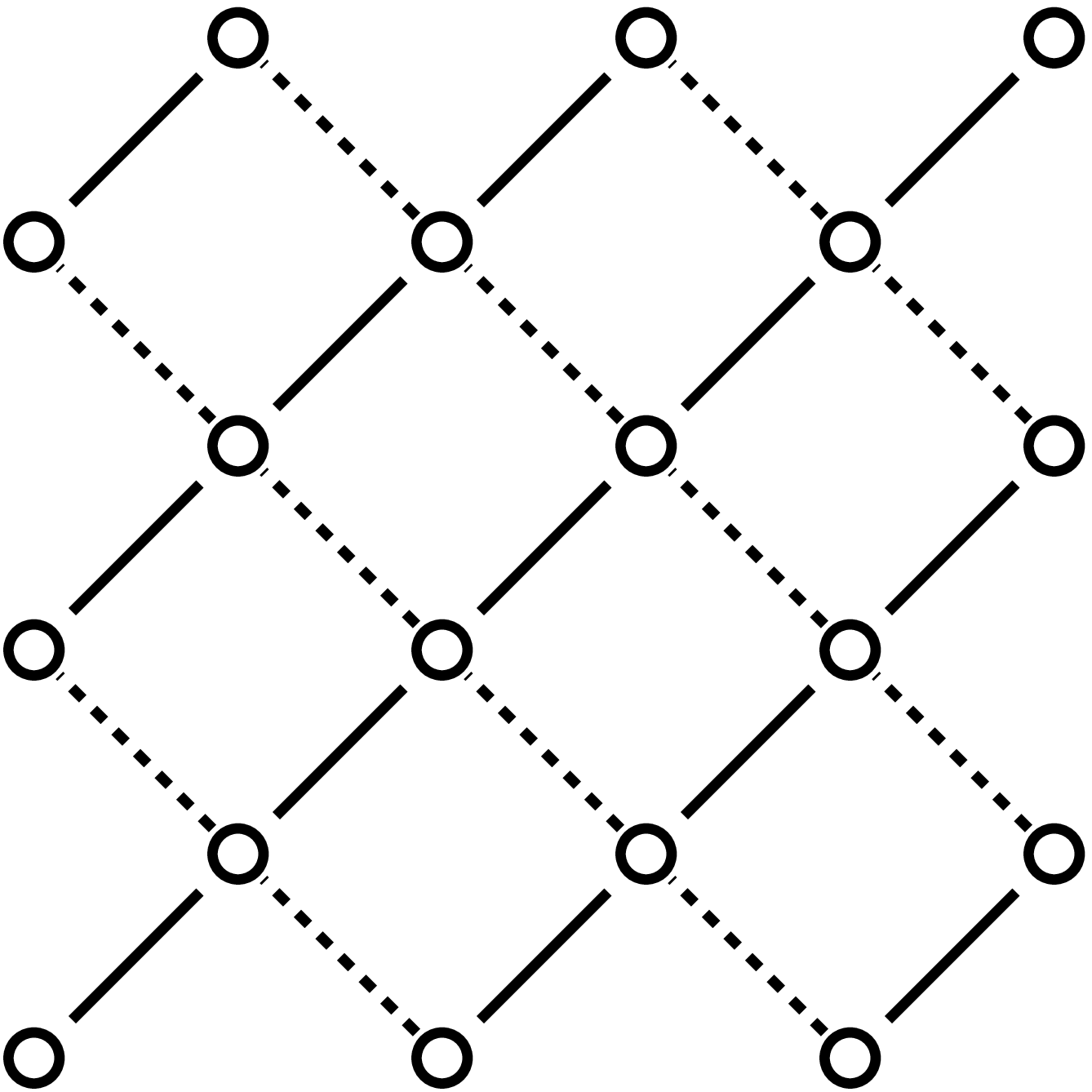}}} 
\put(60,-30){
\epsfxsize=100mm
\epsfysize=40mm
\epsfbox{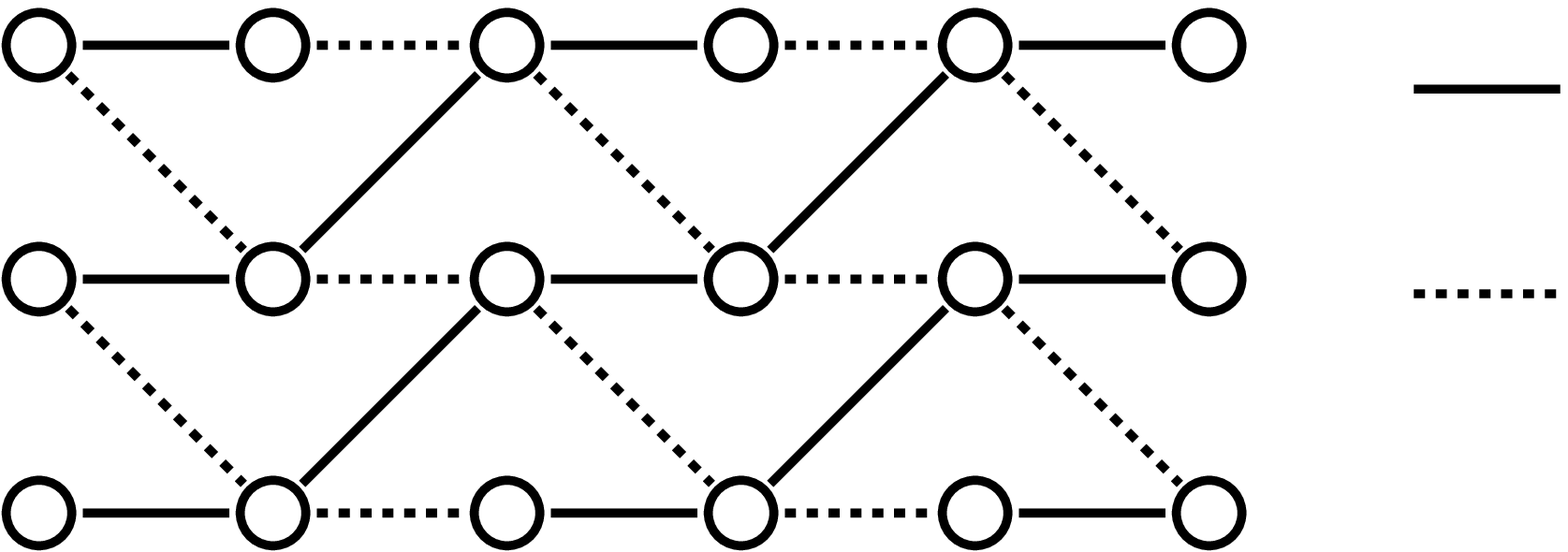}}
\put(20,185){\huge\bf a)}
\put(230,185){\huge\bf b)}
\put(20,22){\huge\bf c)}
\put(290,63){\large\bf ferromagnetic }
\put(290,48){\large\bf bond}
\put(290,22){\large\bf antiferromagnetiic }
\put(290,7){\large\bf bond}
\end{picture}

\vspace{20mm}
\caption{ }
\end{figure}

\clearpage

\begin{figure}[p]
\begin{center}
\begin{picture}(0,0)
\setlength{\unitlength}{1.000pt}
\put(65,+215){\LARGE\bf $L_{v}=1$}
\put(0,+140){\LARGE\bf $L_{v}=2$}
\put(-60, 35){\LARGE\bf $L_{v}=3$}
\end{picture}
\epsfxsize=100mm
\epsfysize=80mm
\epsfbox{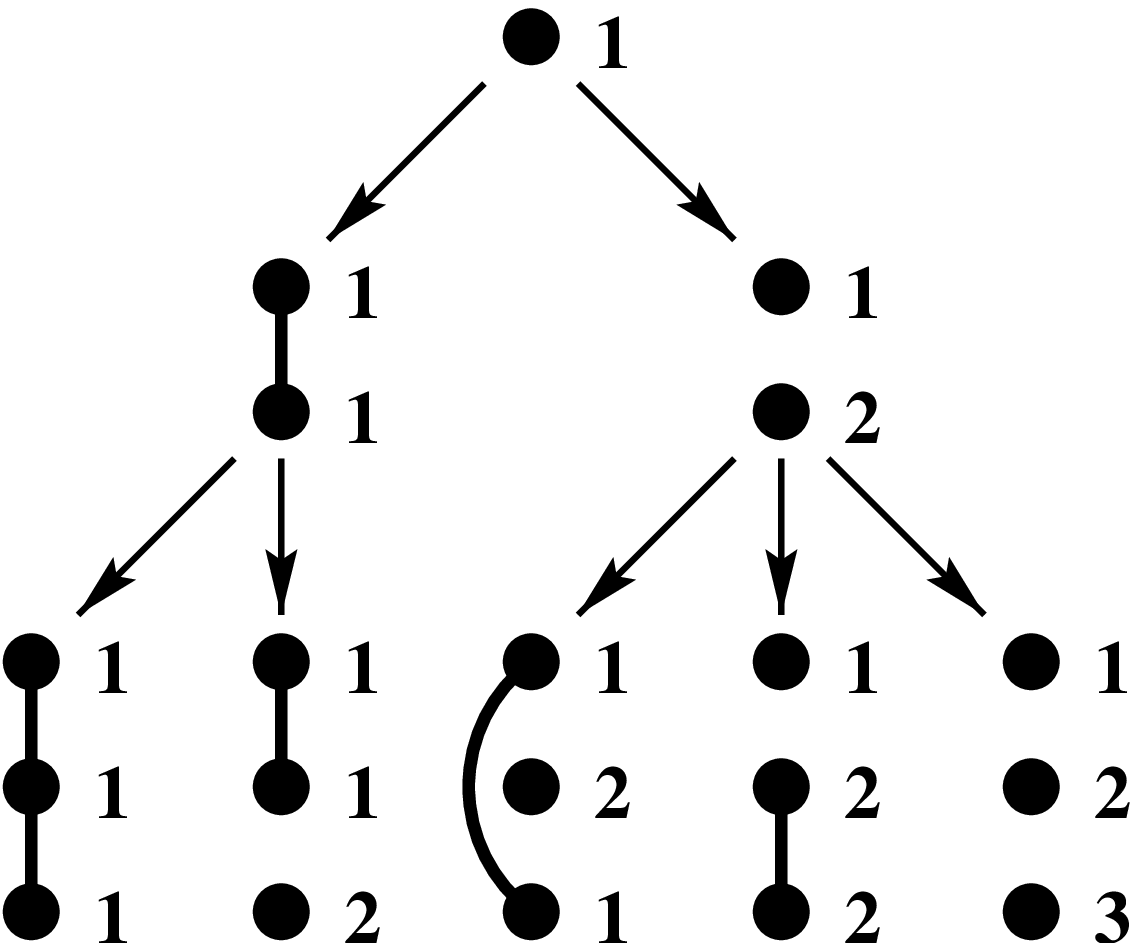}
\end{center}
\caption{ }
\end{figure}

\clearpage

\begin{figure}[p]
\begin{picture}(200,50)
\setlength{\unitlength}{1.000pt}
\put(40,0){
\epsfxsize=90mm
\epsfysize=30mm
\epsfbox{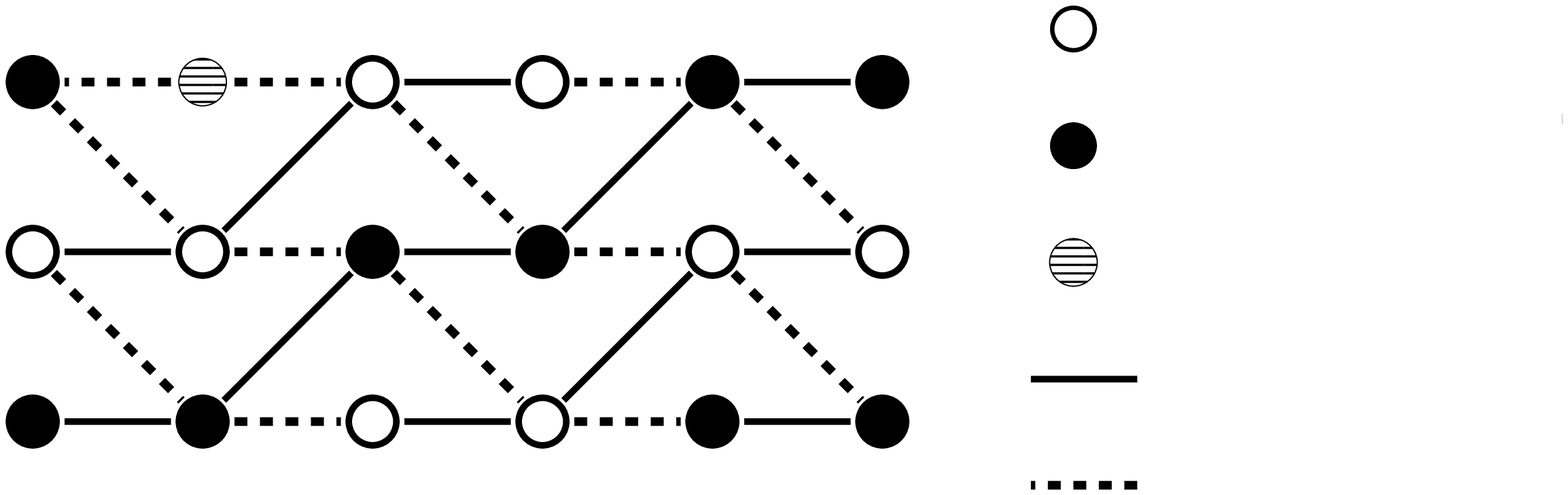}}
\put(40,-110)
{
\epsfxsize=100mm
\epsfysize=30mm
\epsfbox{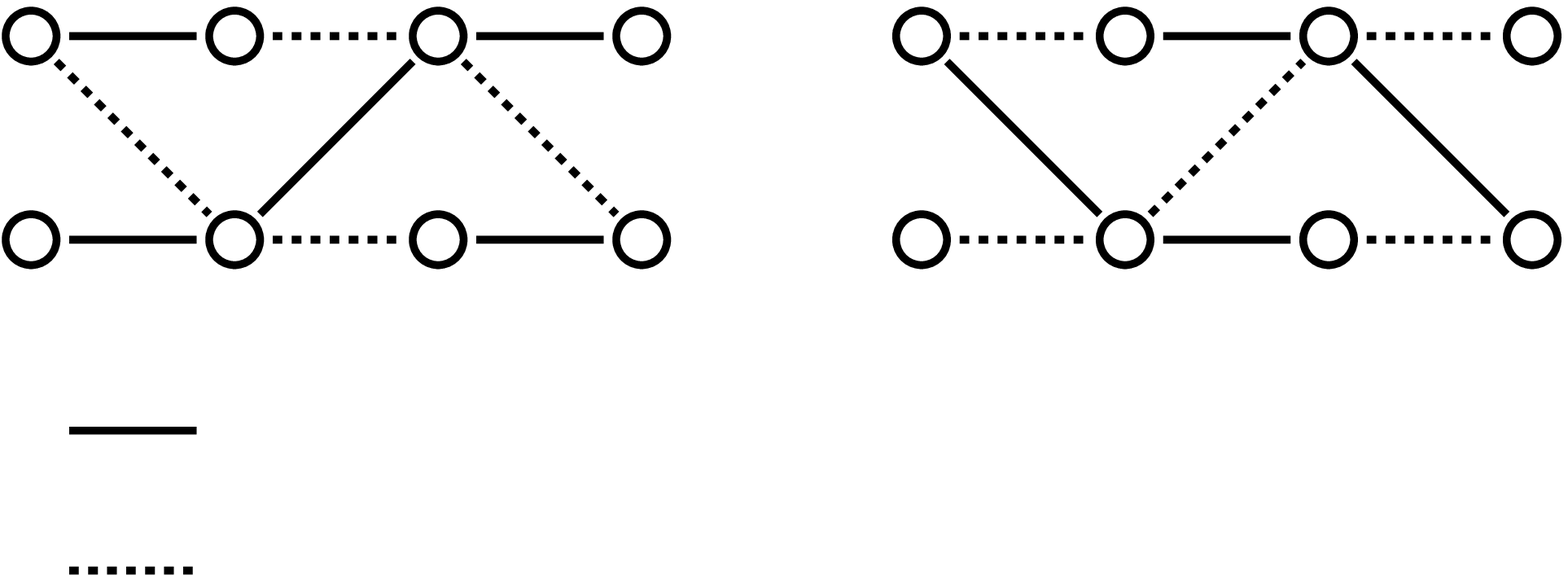}}
\put(40,-200)
{
\epsfxsize=66mm
\epsfysize=20mm
\epsfbox{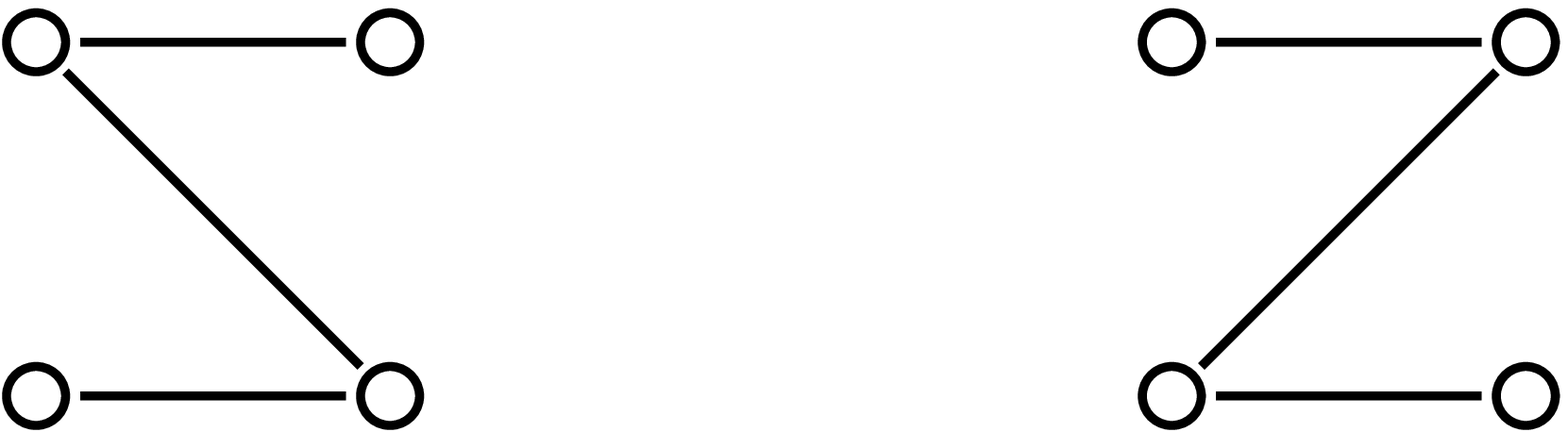}}

\put(60,-144){\large$J_{1,j}$}
\put(60,-205){\large$J_{2,j}$}
\put(67,-171){\large$J_{3,j}$}
\put(18,-150){\Large $\sigma_{1,j}$}
\put(18,-198){\Large $\sigma_{2,j}$}
\put(97,-150){\Large $\sigma_{1,j+1}$}
\put(97,-198){\Large $\sigma_{2,j+1}$}
\put(196,-144){\large$J_{1,j}$}
\put(196,-205){\large$J_{2,j}$}
\put(205,-176){\large$J_{3,j}$}
\put(152,-150){\Large $\sigma_{1,j}$}
\put(152,-198){\Large $\sigma_{2,j}$}
\put(233,-150){\Large $\sigma_{1,j+1}$}
\put(233,-198){\Large $\sigma_{2,j+1}$}
\put(25,90){\large\bf a)}
\put(25,-30){\large\bf b)}
\put(190,-30){\large\bf c)}
\put(25,-135){\large\bf d)}
\put(170,-135){\large\bf e)}
\put(230,17){$J=1$}
\put(230,-1){$J=-1$}
\put(230,80){$\sigma_{i,j}=1$}
\put(230,60){$\sigma_{1,1}=2$}
\put(230,40){$\sigma_{1,2}=3$}
\put(40,82){$\sigma_{1,1}$}
\put(65,82){$\sigma_{1,2}$}
\put(83,-90){\bf ferromagnetic bond}
\put(83,-110){\bf antiferromagnetic bond}

\end{picture}

\vspace{85mm}
\caption{ }
\end{figure}

\clearpage

\begin{figure}[p]
\begin{picture}(200,100)
\setlength{\unitlength}{1.000pt}
\put(-10,-30){
\epsfxsize=150mm
\epsfysize=110mm
\epsfbox{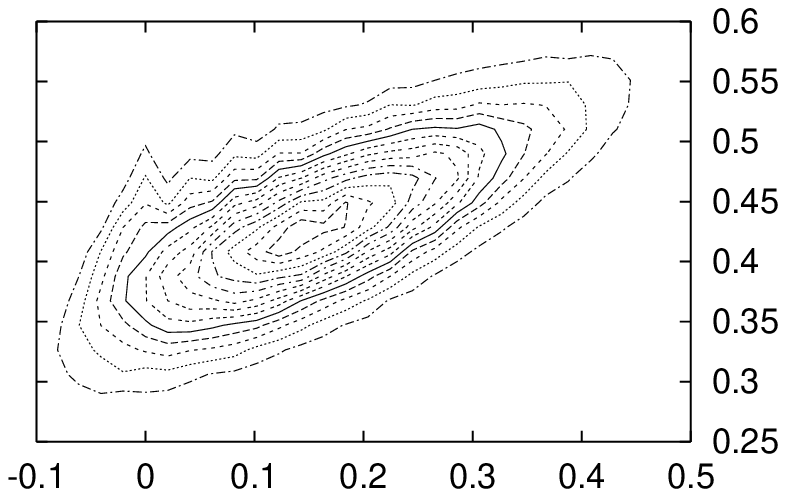}
}
\put(40,-180)
{
\epsfxsize=110mm
\epsfysize=60mm
\epsfbox{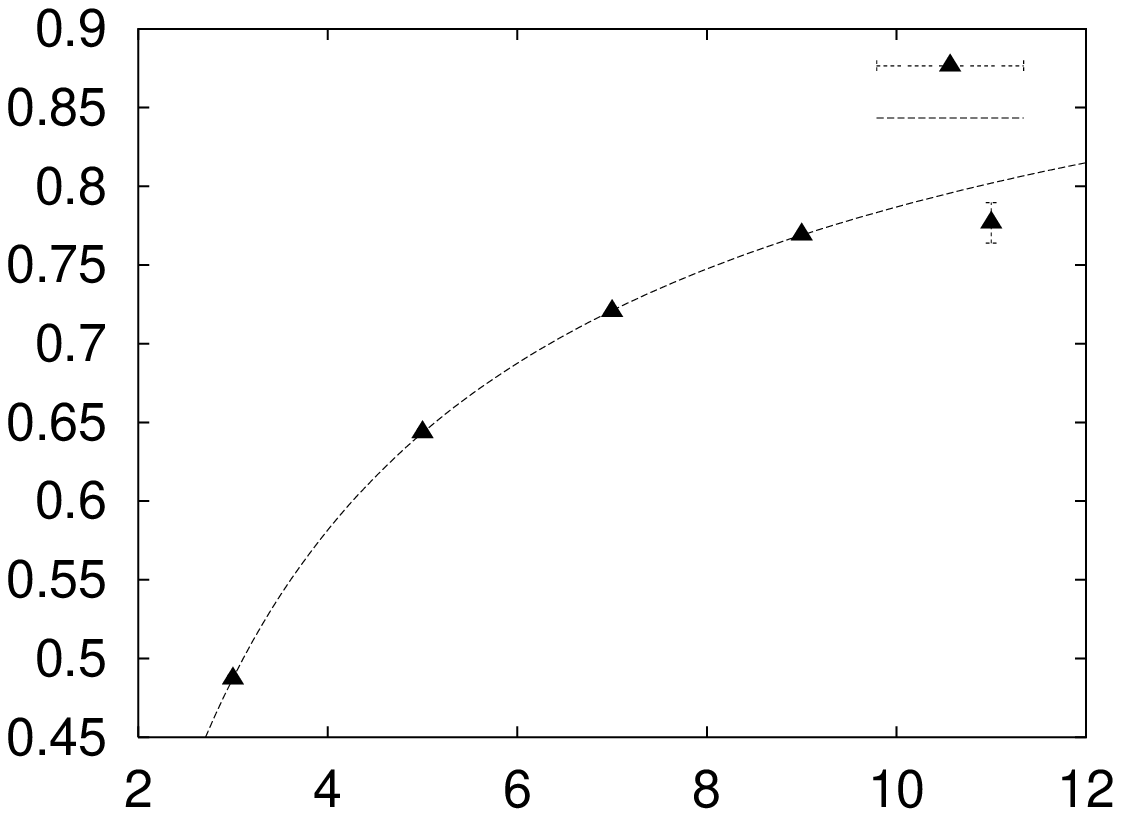}
}
\put(200,25){\Large $\frac{n}{N}$}
\put(60,+136){\large $\frac{|E_{min}|}{N}$}
\put(210,-189){\large $L$}
\put(-5,-85){ ${\rm corr}(n,|E_{min}|)$}
\put(20,140){\Large\bf a)}
\put(20,-40){\Large\bf b)}
\put(180,-31){\footnotesize ${\rm corr}(n,-E_{min})$}
\put(160,-44){\footnotesize $1.04(4)-1.126(14)L^{-0.65(7)}$}
\end{picture}

\vspace{70mm}

\caption{ }
\end{figure}

\clearpage

\begin{figure}[p]
\begin{picture}(200,100)
\setlength{\unitlength}{1.000pt}
\put(40,0){
\epsfxsize=110mm
\epsfysize=60mm
\epsfbox{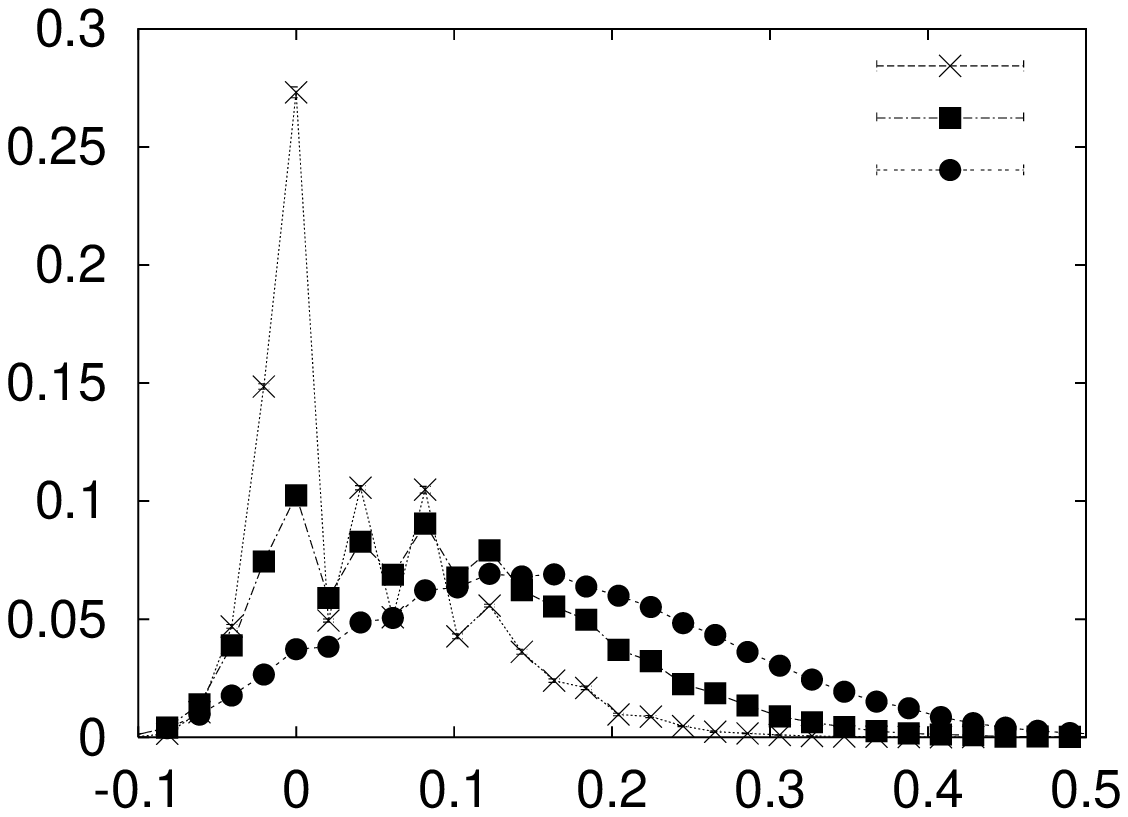}
}
\put(40,-180)
{
\epsfxsize=110mm
\epsfysize=60mm
\epsfbox{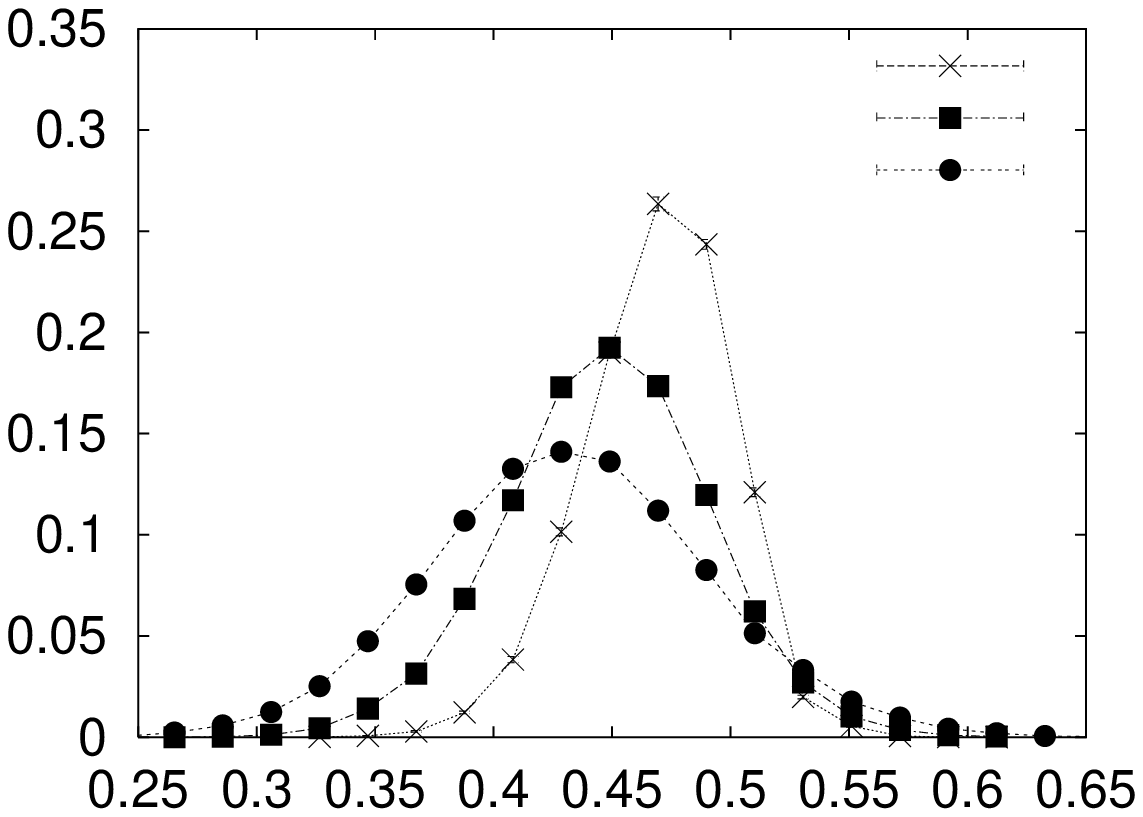}
}
\put(210,-8){\Large $\frac{n}{N}$}
\put(24,+110){\large $P(\frac{n}{N};p)$}
\put(204,-192){\large $\frac{|E_{min}|}{N}$}
\put(5,-83){\large $P(\frac{|E_{min}|}{N};p)$}
\put(20,140){\Large\bf a)}
\put(20,-40){\Large\bf b)}
\put(245,149){\footnotesize $p=0.1$}
\put(245,139){\footnotesize $p=0.2$}
\put(245,128){\footnotesize $p=0.5$}
\put(245,-31){\footnotesize $p=0.1$}
\put(245,-42){\footnotesize $p=0.2$}
\put(245,-52){\footnotesize $p=0.5$}
\end{picture}

\vspace{70mm}
\caption{ }
\end{figure}

\clearpage

\begin{figure}[p]
\begin{picture}(200,100)
\setlength{\unitlength}{1.000pt}
\put(40,0){
\epsfxsize=110mm
\epsfysize=60mm
\epsfbox{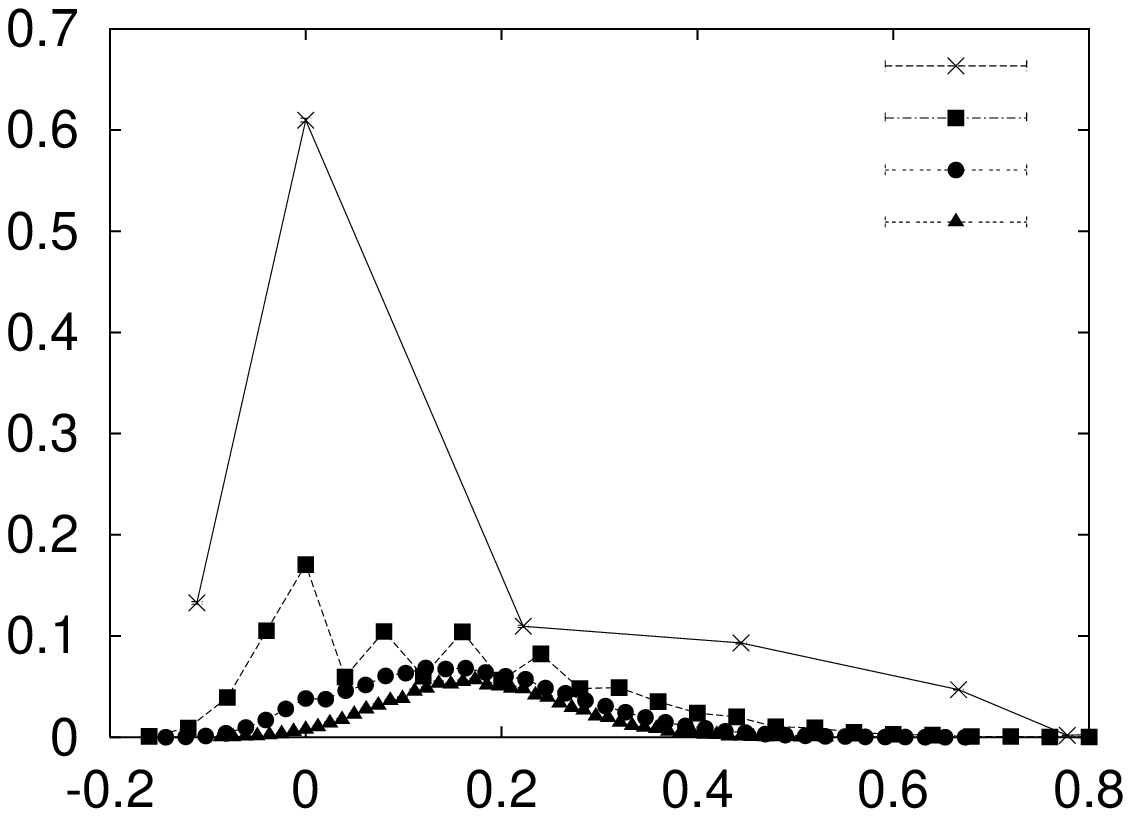}
}
\put(40,-180)
{
\epsfxsize=110mm
\epsfysize=60mm
\epsfbox{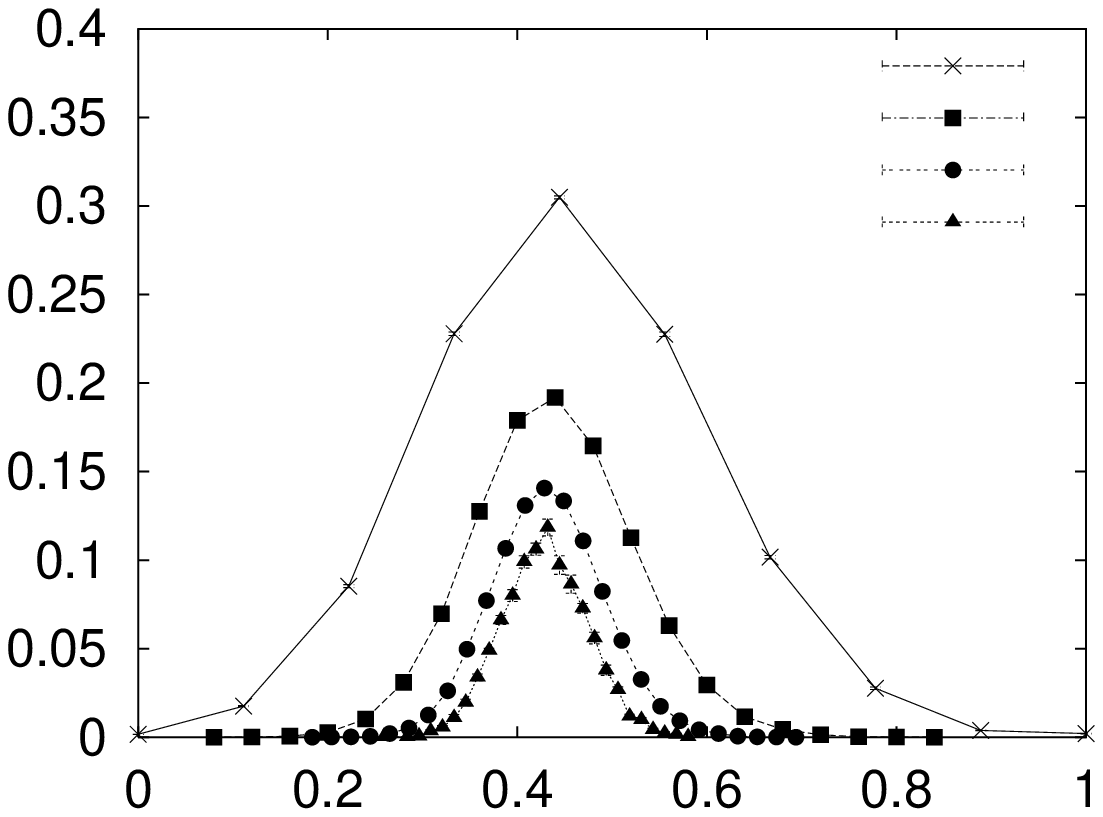}
}
\put(200,-5){\Large $\frac{n}{N}$}
\put(15,+96){\large $P(\frac{n}{N})$}
\put(198,-189){\Large $|\frac{E_{min}}{N}|$}
\put(15,-93){\large $P(|\frac{E_{min}}{N}|)$}
\put(20,140){\Large\bf a)}
\put(20,-40){\Large\bf b)}
\put(250,149){\footnotesize $N=9$}
\put(248,139){\footnotesize $N=25$}
\put(248,128){\footnotesize $N=49$}
\put(248,117){\footnotesize $N=81$}
\put(250,-31){\footnotesize $N=9$}
\put(248,-42){\footnotesize $N=25$}
\put(248,-52){\footnotesize $N=49$}
\put(248,-61){\footnotesize $N=81$}
\end{picture}

\vspace{70mm}

\caption{ }
\end{figure}

\clearpage
\begin{figure}[p]
\begin{picture}(200,100)
\setlength{\unitlength}{1.000pt}
\put(40,0){
\epsfxsize=110mm
\epsfysize=60mm
\epsfbox{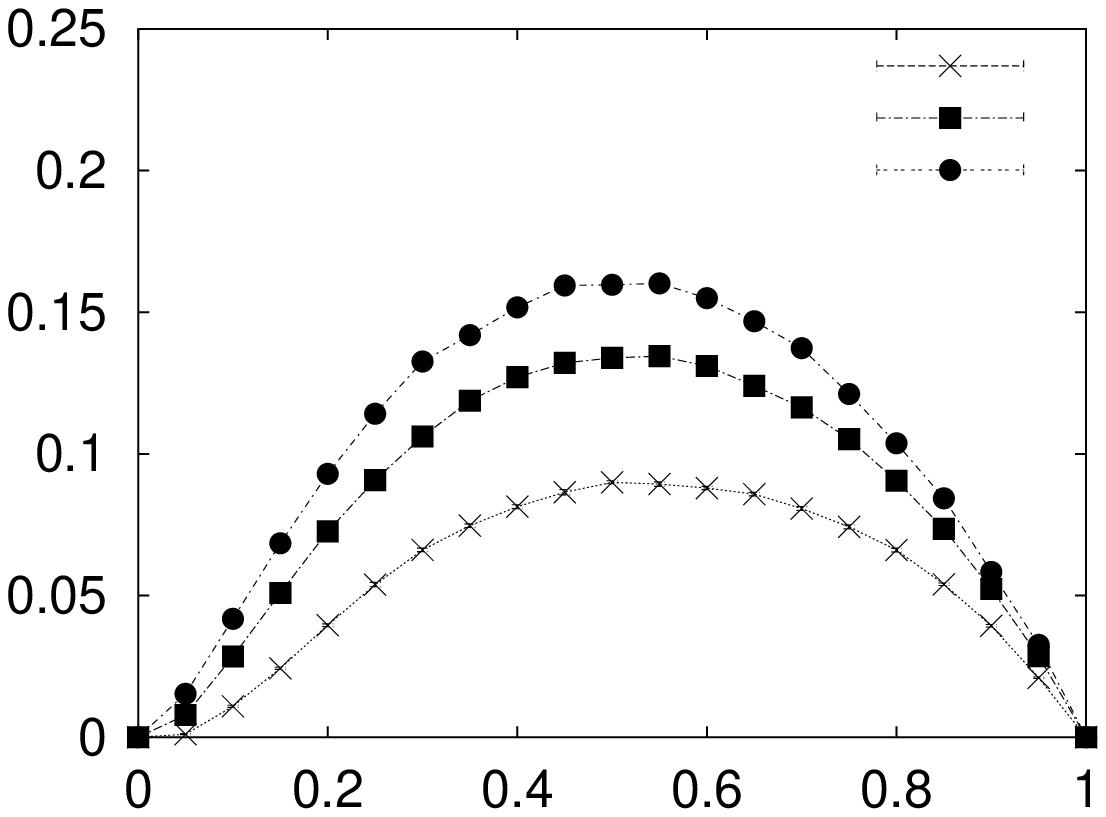}
}
\put(40,-180)
{
\epsfxsize=110mm
\epsfysize=60mm
\epsfbox{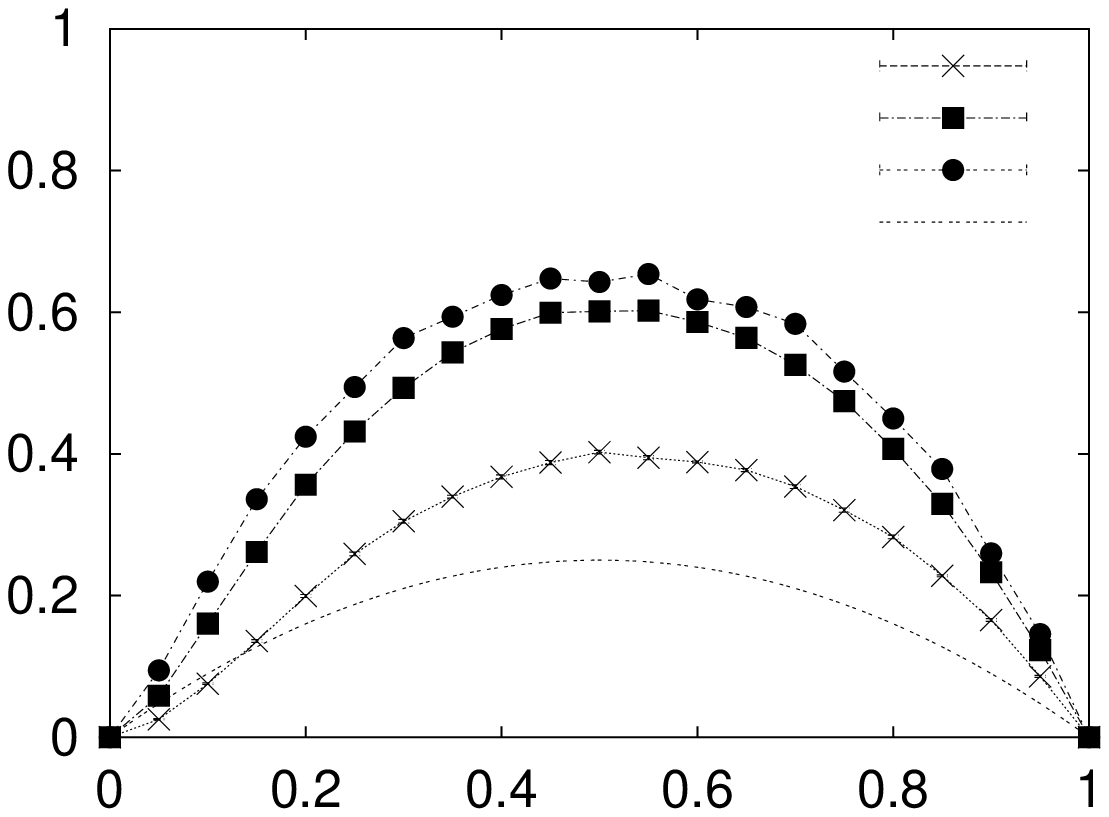}
}
\put(210,-5){\Large $p$}
\put(35,+90){\large $\left< \frac{n}{N} \right>$}
\put(210,-180){\Large $p$}
\put(25,-90){\large $W_{knot}^{2}$}
\put(20,140){\Large\bf a)}
\put(20,-40){\Large\bf b)}
\put(252,149){\footnotesize $N=9$}
\put(248,139){\footnotesize $N=25$}
\put(248,128){\footnotesize $N=49$}
\put(250,-31){\footnotesize $N=9$}
\put(248,-42){\footnotesize $N=25$}
\put(248,-52){\footnotesize $N=49$}
\put(241,-61){\scriptsize $0.5 p(1-p)$}
\end{picture}

\vspace{70mm}

\caption{ }
\end{figure}

\clearpage
\begin{figure}[p]
\begin{picture}(200,100)
\setlength{\unitlength}{1.000pt}
\put(40,0){
\epsfxsize=110mm
\epsfysize=60mm
\epsfbox{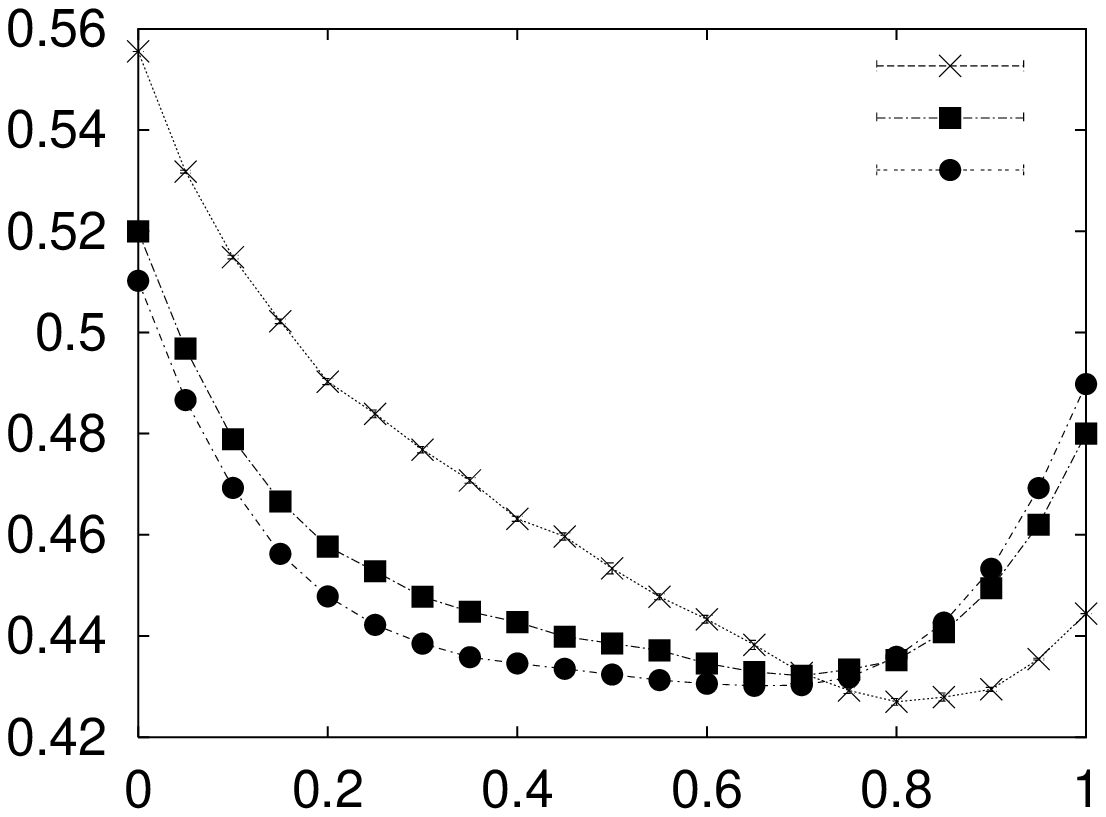}
}
\put(40,-180)
{
\epsfxsize=110mm
\epsfysize=60mm
\epsfbox{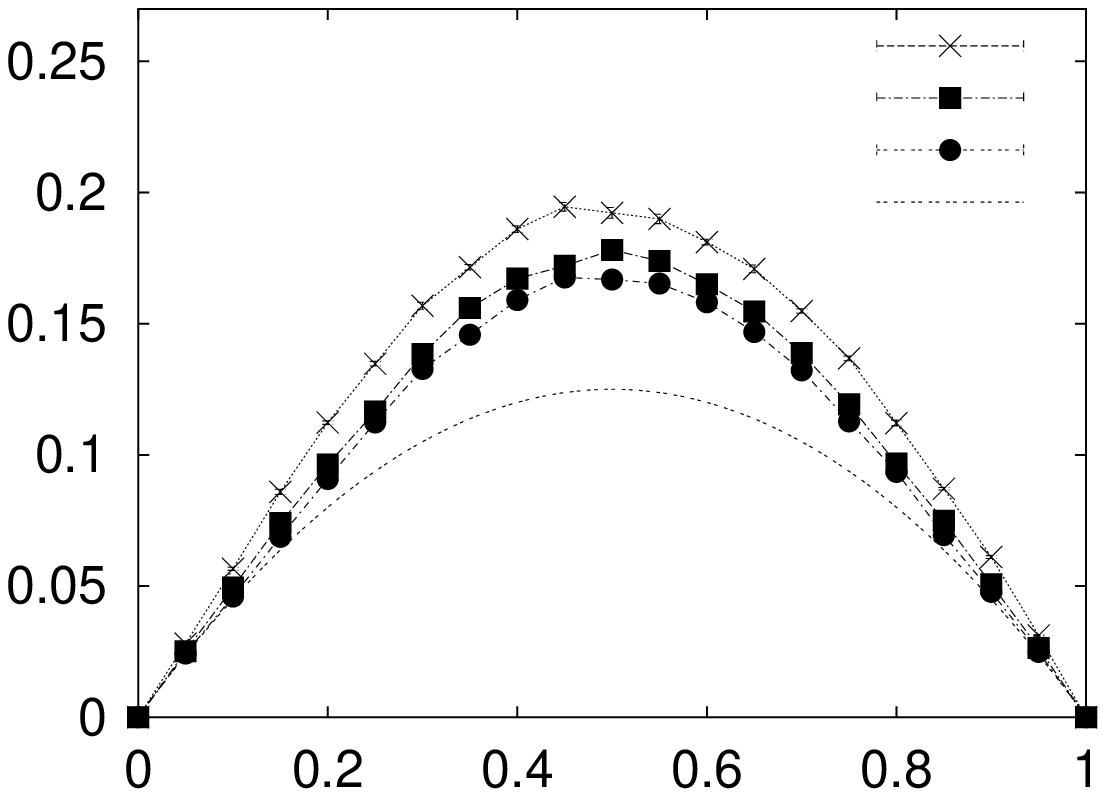}
}
\put(210,-5){\Large $p$}
\put(20,+96){\large $\left< \frac{|E_{min}|}{N} \right>$}
\put(210,-180){\Large $p$}
\put(25,-90){\large $W_{Potts}^{2}$}
\put(20,140){\Large\bf a)}
\put(20,-40){\Large\bf b)}
\put(250,149){\footnotesize $N=9$}
\put(248,139){\footnotesize $N=25$}
\put(248,128){\footnotesize $N=49$}
\put(250,-31){\footnotesize $N=9$}
\put(248,-42){\footnotesize $N=25$}
\put(248,-52){\footnotesize $N=49$}
\put(241,-61){\scriptsize $0.5 p(1-p)$}
\end{picture}

\vspace{70mm}

\caption{ }
\end{figure}

\clearpage 

\begin{figure}[p]
\begin{picture}(200,100)
\setlength{\unitlength}{1.000pt}
\put(40,0){
\epsfxsize=110mm
\epsfysize=60mm
\epsfbox{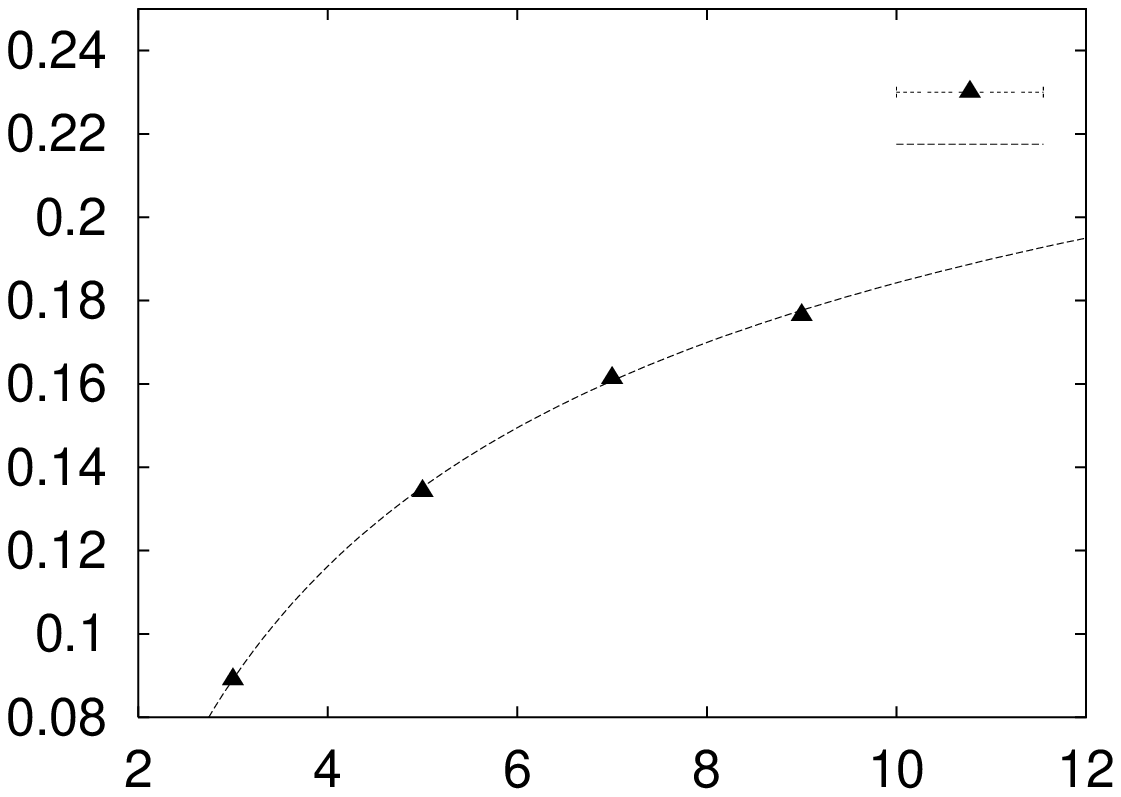}
}
\put(40,-180)
{
\epsfxsize=110mm
\epsfysize=60mm
\epsfbox{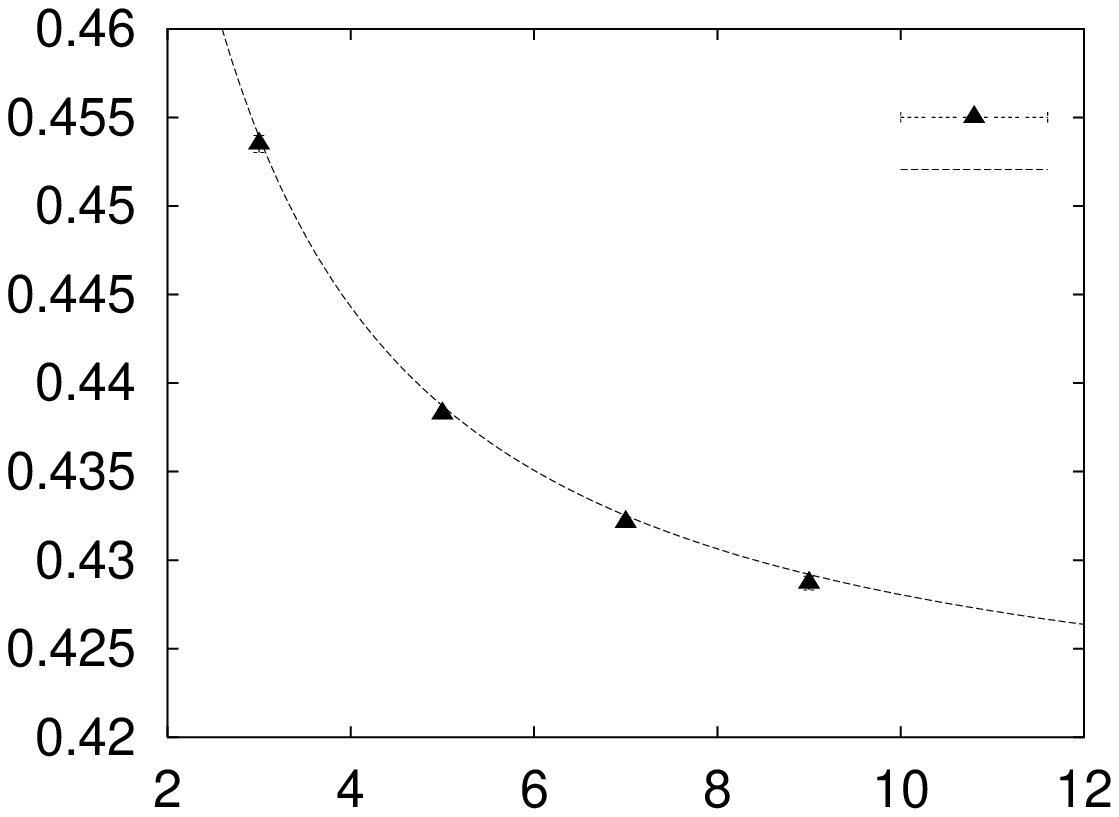}
}
\put(208,-5){\large $L$}
\put(30,+105){\large $\left<\frac{n}{N} \right>$}
\put(208,-185){\large $L$}
\put(10,-80){ \large $ \left< \frac{|E_{min}|}{N} \right>$}
\put(20,140){\Large\bf a)}
\put(20,-40){\Large\bf b)}
\put(224,141){\footnotesize  $<n/N>$}
\put(170,127){\footnotesize $0.334(8)-0.41(2)L^{-0.38(5)}$}
\put(224,-40){\footnotesize $< |E_{min}|/N >$}
\put(165,-54){\footnotesize $0.4185(7)+0.119(3)L^{-1.1(4)} $}
\end{picture}

\vspace{70mm}
\caption{ }
\end{figure}

\clearpage

}


\begin{figure}[p]
\begin{picture}(200,300)
\setlength{\unitlength}{1.000pt}
\put(40,0){
\epsfxsize=110mm
\epsfysize=60mm
\epsfbox{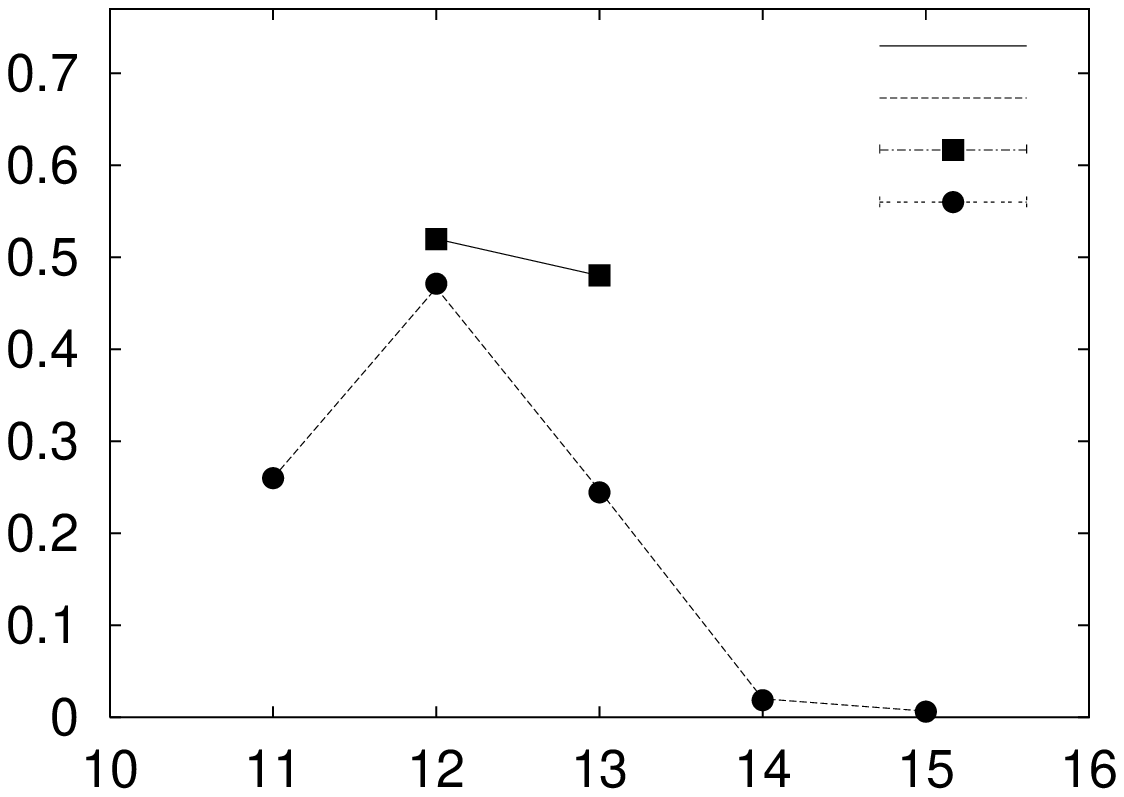}
}
\put(40,-180)
{
\epsfxsize=110mm
\epsfysize=60mm
\epsfbox{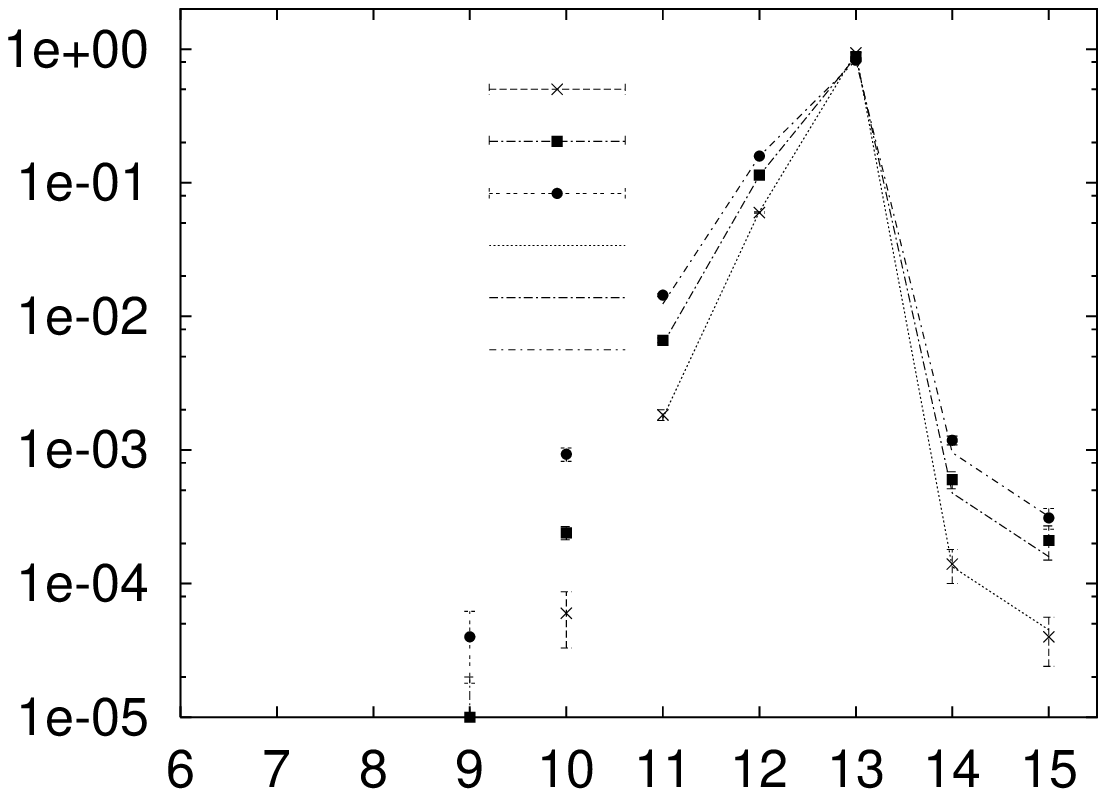}
}

\put(212,+150){\footnotesize analyt. $M=1$}
\put(212,+140){\footnotesize  analyt. $M=2$}
\put(205,+130){\footnotesize numeric. $M=1$}
\put(205,+119){\footnotesize numeric. $M=2$}
\put(120,-39){\footnotesize num.$p=0.01$}
\put(120,-49){\footnotesize num.$p=0.02$}
\put(120,-59){\footnotesize num.$p=0.03$}
\put(112,-69){\footnotesize analyt.$p=0.01$}
\put(112,-79){\footnotesize analyt.$p=0.02$}
\put(112,-89){\footnotesize analyt.$p=0.03$}
\put(5,+90){\large $P(|E_{min}|)$}
\put(8,-92){\large $P(|E_{min}|)$}
\put(200,-189){\large $|E_{min}|$}
\put(200,-7){\large $|E_{min}|$}
\put(20,140){\Large\bf a)}
\put(20,-43){\Large\bf b)}
\end{picture}

\vspace{75mm}
\caption{ } 
\end{figure}

\end{document}